\pdfoutput=1
\documentclass[aps,prd,preprint,groupedaddress,showpacs,showkeys,nofootinbib]{revtex4-1}

\usepackage{graphicx}
\usepackage{dcolumn}
\usepackage{bm}
\usepackage{amsmath}
\usepackage{amssymb} 
\usepackage{mathrsfs}
\usepackage{tensor}
\usepackage{slashed}
\usepackage{feynmf}
\usepackage[dvipsnames]{xcolor}
\usepackage[colorlinks]{hyperref}
\hypersetup{linktocpage}
\usepackage{multirow}
\usepackage{latexsym}
\usepackage{graphicx}
\usepackage[utf8]{inputenc}
\usepackage{verbatim}
\usepackage{colortbl}
\usepackage{fancyhdr}
\usepackage{caption}
\usepackage{subcaption}
\usepackage[font=footnotesize,labelfont=bf,justification=justified]{caption}
\newcommand{\dbar}{d\hspace*{-0.08em}\bar{}\hspace*{0.1em}}

\begin{document}
\-\vspace{3cm}
\title{Non-Gaussianities in a two-field generalization of Natural Inflation.}

\author{Simon Riquelme M.}
\email{sdriquel@umd.edu}
\affiliation{Maryland Center for Fundamental Physics, Department of Physics, University of Maryland, College Park, MD 20742-4111}


\begin{abstract}
We describe a two-field model that generalizes Natural Inflation, in which the inflaton is the pseudo-Goldstone boson of an approximate symmetry that is spontaneously broken, and the radial mode is
dynamical. We analyze how the dynamics fundamentally depends on the mass of the radial mode and calculate/estimate the non-Gaussianities arising from such a scenario.
\end{abstract}


\maketitle

\tableofcontents

\newpage

\section{Introduction.}\label{intro}

Inflation is a well established framework that resolves several puzzles in big bang cosmology. The well known flatness, horizon and monopole problems can successfully be tackled by demanding a period of quasi-exponential expansion of the early universe \cite{Guth:1980zm,Linde:1981mu,Albrecht:1982wi}. While this classical picture is quite nice by itself, the quantum implications of this idea are also far reaching. Roughly speaking, all the structure in the universe can be understood as arising from primordial quantum fluctuations of the inflaton field \cite{Mukhanov:1981xt}. To successfully match the percent-level deviation from perfect scale-invariance of the power spectrum of gauge-invariant primordial curvature perturbations that current observations demand, a considerable exponential growth of the scale factor $a = a(t)$ of the (flat) Friedmann-Lema\^itre-Robertson-Walker (FLRW) geometry, with metric $ds^2 = -dt^2 + a^2(t)\,\bold{dx^2}$, is required, which is translated into several ``slow-roll'' conditions over the potential for the inflaton field. Many models that realize this slow-roll inflation scenario have been proposed over the years \cite{Martin:2013tda,Lyth:2009zz}.
\\
\indent The power spectrum contains all the information about the primordial perturbations \textit{if} the initial conditions are drawn from a Gaussian distribution function. However, higher-order correlations may encode a significant amount of new information, as they are sensitive to non-linear interactions, while the power spectrum only probes the free theory. In the early 2000s, Maldacena proved that so-called non-Gaussianities for primordial scalar fluctuations in the simplest \footnote{By simplest we mean single scalar field slow-roll inflation with a canonical kinetic term plus Einstein gravity using the so-called Bunch-Davies vacuum.} inflationary models are generically suppressed by slow-roll parameters \cite{Maldacena:2002vr}, meaning $f_\text{NL} \sim \mathcal{O}(\epsilon, \eta)$, where the non-linear parameter $f_\text{NL}$ is a measure of the amplitude of non-Gaussianities, $\epsilon \equiv -\frac{\dot{H}}{H^2}$ and $\eta \equiv \frac{\dot{\epsilon}}{\epsilon\,H}$ are the usual slow-roll parameters of inflation and the Hubble function is defined as $H \equiv \frac{\dot{a}}{a}$ \footnote{It is now understood that in the local subcase, $f_\text{NL}^\text{local} = 0$, for \textit{all} single-field inflation models \cite{Tanaka:2011aj,Pajer:2013ana} (we thank G. Palma for pointing this out to us). See appendix \ref{nongaussdef} for an exact definition of $f_\text{NL}^\text{local}$.}. Consequently, since slow-roll conditions demand $\{\epsilon,\eta\} \ll 1$, we should abandon the possibility of observing such features if Nature really picked up this single-field slow-roll scenario as it is highly unlikely that we will ever be able to disentangle these ``quantum'' non-Gaussianities from ``classical'' ones that arise from CMB evolution \cite{Bartolo:2010qu} and from LSS \cite{2010AdAst2010E..73L} (due to the non-linear gravitational evolution or the galaxy bias), with $f_\text{NL} \sim \mathcal{O}(1)$ as the natural size of these effects.
\\
\indent One way out of this ``no-go'' situation is to consider the so-called $P(X)$ theories \cite{ArmendarizPicon:1999rj,Garriga:1999vw}, where $X \equiv -\frac{1}{2}(\partial_\mu\varphi)^2$ and $\varphi$ denotes the inflaton field. These theories may produce large non-Gaussianities without disrupting the inflationary background solution by respecting a mildly broken shift symmetry $\varphi \to \varphi + \text{constant}$, though it is important to keep in mind that it is a challenge to find a radiatively stable $P(X)$ scenario \footnote{One example of a radiatively stable UV-completion, where the form of the action is protected by a ``higher-dimensional boost symmetry'' is the case of DBI inflation \cite{Silverstein:2003hf,Alishahiha:2004eh}.}. It has been found that $P(X)$ theories generically predict that $f_\text{NL} \sim c_s^{-2}$, where $c_s$ is the ``speed of sound'' of adiabatic fluctuations. Consequently, in principle, a non-trivial (small) speed of sound can lead to observable non-Gaussianities. 
\\
\indent Another logical possibility is to consider additional fields during inflation. One crucial property of these fields is their mass, collectively denoted as $m$, compared to the Hubble scale $H$. There is an extensive literature regarding the case when these extra fields are light or even massless so that $m^2 \ll H^2$ (see \cite{Wands:2007bd,Langlois:2012sz} for a review). This range of masses implies that non-Gaussianities will be effectively generated from non-linearities \textit{after} horizon crossing, when all modes have become classical. At the other end, in the very massive case, meaning $m^2 \gg H^2$, we can always ``integrate out'' the heavy fields, leading to a simplified theory by producing new (non-slow-roll) operators in the effective field theory (EFT) for the inflaton. The so-called Quasi-Single-Field (QSF) inflation models \cite{Chen:2009we,Chen:2009zp,Baumann:2011nk,Assassi:2012zq,Chen:2012ge,Noumi:2012vr,Arkani-Hamed:2015bza,Gong:2013sma,Emami:2013lma,Kehagias:2015jha,Chen:2015lza,Bonga:2015urq,Lee:2016vti,Chen:2017ryl} explore the third relevant regime, $m^2 \sim H^2$, where the new particles can in principle be produced by quantum fluctuations during the inflationary stage and then decay into inflatons, leaving a statistical imprint on the spectrum of primordial fluctuations. Importantly, the production of these particles gives rise to non-local effects which cannot be captured by a single-field EFT and can potentially give rise to observable non-Gaussianities. There are several arguments for why it is reasonable to expect that the inflationary paradigm should naturally incorporate particles with such masses \footnote{For example, consider the case when supersymmetry (SUSY) is invoked to tame the quantum corrections to the inflationary potential. Under the assumption that SUSY is not broken at energies higher than the inflationary scale $H$, the vacuum energy during inflation will surely break it as there is no supersymmetric theory in de Sitter space. This implies that additional fields which are not protected by global symmetries will inherit Hubble scale masses from SUSY breaking (this is related to the so-called ``eta problem'' of supergravity inflation models \cite{Copeland:1994vg}). See \cite{Baumann:2011nk} for details.} and how they may show up in the ``cosmological experiment'', as has been recently emphasized in \cite{Arkani-Hamed:2015bza}.

\newpage

\indent In this paper, we will introduce and explore a two-field model that we unimaginatively dub ``Generalized Natural Inflation'' (GNI), a well-motivated generalization or ``UV-completion'' of the influential Natural Inflation (NI) scenario \cite{Freese:1990rb,Adams:1992bn}. Let us recall that single-field NI originally conceived the seminal idea that the inflaton is a \textit{pseudo}-Nambu-Goldstone boson (pNGB) so it naturally has an exceptionally flat potential, which is a slow-roll requirement. In our model the inflaton plays the role of the phase $\theta$ of a complex scalar field $\chi \sim \sigma\,e^{i\theta}$, and the radial mode $\sigma$ is taken to be dynamical, with a mass $m_\sigma$ determined by the spontaneous breaking of a global $U(1)$ symmetry. To give a small mass to the would-be Goldstone (inflaton) field, so slow-roll conditions are satisfied, we softly break the $U(1)$ symmetry by a relevant operator. We will consider the cases $m_\sigma^2 \gg H^2$ and $m_\sigma^2 \sim H^2$ and find estimates for the non-Gaussianities that may arise in these scenarios \footnote{We briefly consider the case $m_\sigma^2 \ll H^2$ in subsection \ref{commlightmass}, where we demonstrate why this case is rather uninteresting for our particular model.}. The latter QSF regime is specially interesting, as we are effectively able to constrain an a priori arbitrary potential for the so-called ``isocurvature'' mode of the original (vanilla) QSF model of Chen and Wang \cite{Chen:2009we,Chen:2009zp}.
\\
\indent This paper is structured as follows. In section \ref{infback} we introduce our model and go through the analysis of its associated inflationary background solution. We discuss how suitable initial conditions can lead to observable non-Gaussianities by dynamically decreasing the speed of sound of adiabatic fluctuations. We calculate the observables of the inflationary model and discuss its current viability given updated bounds coming from Planck 2015 \cite{Ade:2015xua} and Planck/Bicep \cite{Array:2015xqh} missions. In section \ref{infpert} we discuss the theory of inflationary perturbations. First, we analyze the case when the radial field is very massive so it can be naively integrated out. We contrast the predictions for non-Gaussianities of our model when neglecting \cite{Achucarro:2012sm}, as opossed to taking into account \cite{Gong:2013sma}, the self-interactions of the heavy field. Then we address the QSF regime and obtain quantitative estimates for the size of non-Gaussianities. We find that, contrary to naive expectations, due to the tight observational constraints on the parameters of the model, non-Gaussianities are unobservably small. We conclude in section \ref{discandconc} leaving some technical details for appendices \ref{eftofback}, \ref{eftofinf}, \ref{nongaussdef} and \ref{efftheory}.

\newpage

\section{Generalized Natural Inflation.}\label{infback}

\subsection{Multifield Inflation.} 
Let a ``multifield'' theory \footnote{Usually the so-called multifield inflation scenario is understood to be one equipped with a shift symmetry, i.e., $\sigma^i \to \sigma^i + \text{constant}$ for the non-adiabatic (isocurvature) directions $\sigma^i$, so they remain light \cite{Senatore:2010wk}. We do not assume such a constraint in the present multifield formalism.} be described by the following action \cite{Achucarro:2010da}
\begin{align}
S[g,\phi] = \int d^4x\sqrt{-g}\left(\frac{M_\text{Pl}^2}{2}\,\mathscr{R} - \frac{1}{2}\,\tensor{\gamma}{^a^b}(\phi)\tensor{g}{^\mu^\nu}\partial_\mu\phi_a\partial_\nu\phi_b - V(\phi)\right),\label{multifieldaction}
\end{align}
where $\mathscr{R}$ is the Ricci scalar constructed out of the spacetime metric $\tensor{g}{_\mu_\nu}$, $\gamma(\phi)$ is the ``field metric'' and $\phi^a$ is a ``vector'' in field space. $V(\phi)$ is some potential for the scalar fields and $M_\text{Pl}^2 \equiv (8\,\pi\,G_N)^{-1} \approx (2.43\times10^{18}\,\text{GeV})^2$ where $G_N$ is Newton's constant. From $\gamma(\phi)$ we can construct a Christoffel symbol 
\begin{align}
\tensor{\Gamma}{^a_b_c} = \frac{1}{2}\tensor{\gamma}{^a^d}\left(\partial_b\tensor{\gamma}{_c_d} + \partial_c\tensor{\gamma}{_b_d} - \partial_d\tensor{\gamma}{_b_c}\right),
\end{align}
and a corresponding curvature tensor 
\begin{align}
\tensor{\mathbb{R}}{^a_b_c_d} = \partial_c\tensor{\Gamma}{^a_b_d} - \partial_d\tensor{\Gamma}{^a_b_c} + \tensor{\Gamma}{^a_c_e}\tensor{\Gamma}{^e_d_b} - \tensor{\Gamma}{^a_d_e}\tensor{\Gamma}{^e_c_b}.
\end{align}
Varying \eqref{multifieldaction} with respect to $\phi_a$ we get the field equations
\begin{align}
\square\,\phi^a + \tensor{\Gamma}{^a_b_c}\,\tensor{g}{^\mu^\nu}\partial_\mu\phi^b\partial_\nu\phi^c = V^a,\label{fieldeq}
\end{align}
where $V^a \equiv \tensor{\gamma}{^a^b}\,\partial_bV$. It is amusing to note the resemblance of this set of equations with the geodesic equation of motion of a relativistic particle in a non-trivial spacetime background under the influence of external (non-gravitational) forces.\\
Now if we assume that $\phi^a = \phi^a(t)$ and $ds^2 = -dt^2 + a^2(t)\,\bold{dx}^2$, the field equations for $\phi^a$ along with Einstein's equation for the spacetime metric read
\begin{align}
\frac{D}{dt}\,\dot{\phi}^a + 3H\dot{\phi}^a + V^a &= 0,\label{covfieldeq}\\
H^2 - \frac{1}{3M_\text{Pl}^2}\left(\frac{1}{2}\,\dot{\phi}^2 + V\right) &= 0,\label{fried}\\
\dot{H} + \frac{\dot{\phi}^2}{2M_\text{Pl}^2} &= 0,\label{cons}
\end{align}
where $H \equiv \frac{\dot{a}}{a}$ is the Hubble parameter, $DX^a \equiv dX^a + \tensor{\Gamma}{^a_b_c}\,X^b\,d\phi^c$ is a field space covariant derivative and $\dot{\phi}^2 \equiv \tensor{\gamma}{_a_b}\,\dot{\phi}^a\dot{\phi}^b$ is the squared norm of $\dot{\phi}^a$. It is easy to show that \eqref{cons} is not independent but can actually be derived from \eqref{covfieldeq} and \eqref{fried}.

\subsection{The Model.}
\begin{figure}[h!]
  \begin{subfigure}[b]{0.65\textwidth}
    \includegraphics[width=\textwidth]{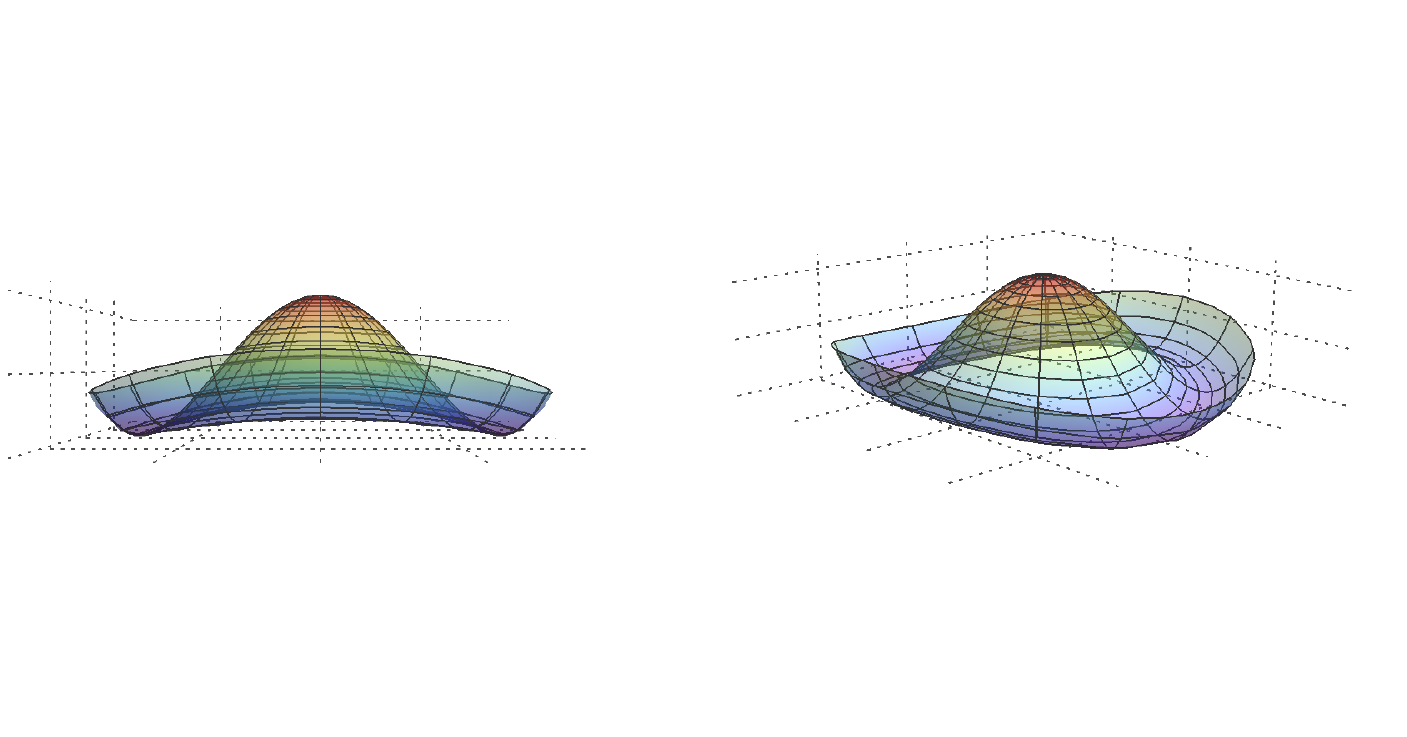}
    \caption{Front and aerial views of $V(\sigma,\theta)$.\\
    We see that the brim of the hat has sinusoidal behavior due to the explicit symmetry breaking.}
    \label{fig:f1}
  \end{subfigure}
  \hfill
  \begin{subfigure}[b]{0.3\textwidth}
    \includegraphics[width=\textwidth]{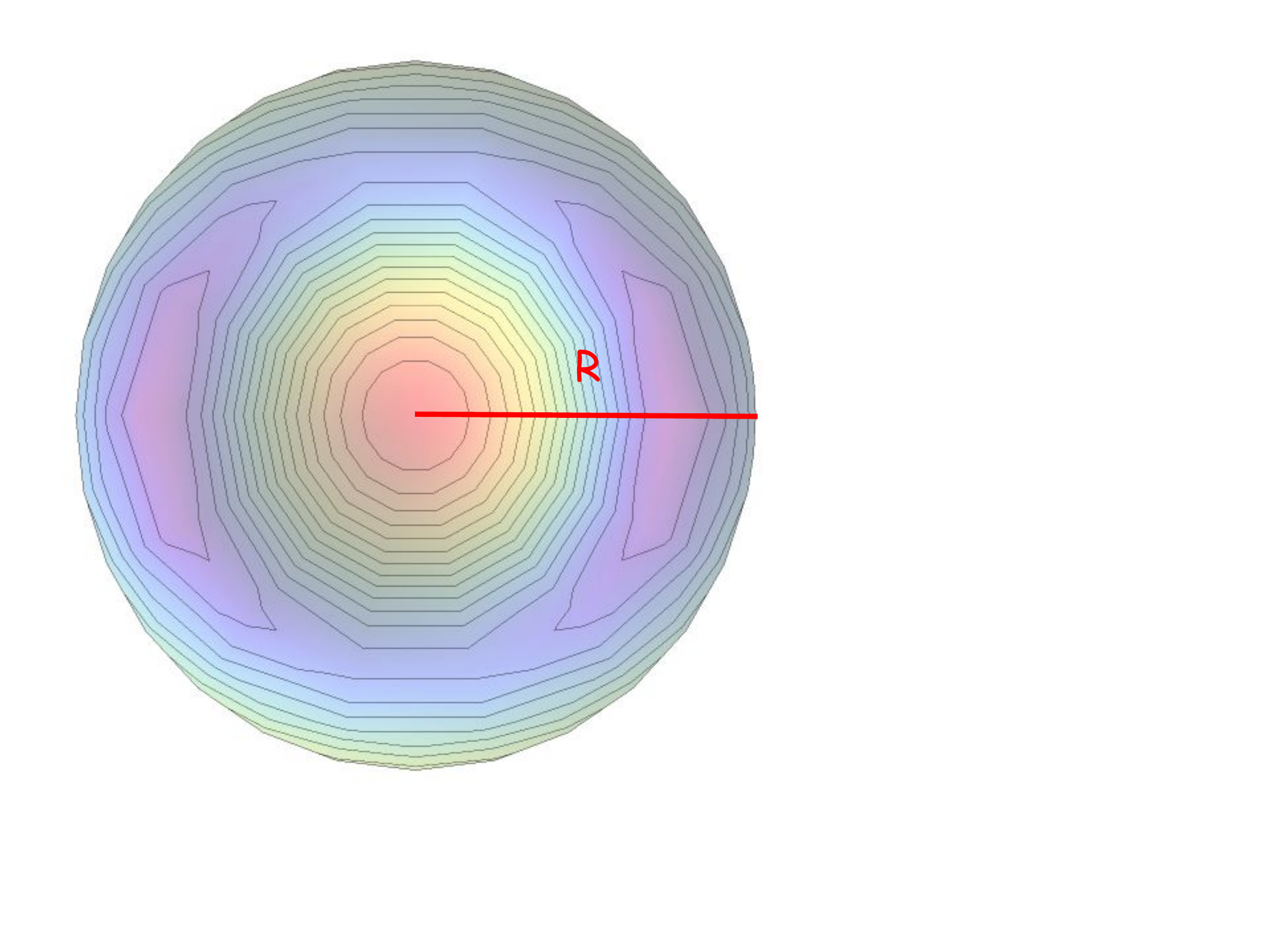}
    \caption{Top view of $V(\sigma,\theta)$.\\
    We see that the contour lines are ellipses and there are different extrema with different radii.}
    \label{fig:f2}
  \end{subfigure}
  \caption{Deformed Mexican Hat.}
  \label{fig:defpot}  
\end{figure}

The model we want to introduce is motivated by the idea that the inflaton field can be identified as the pseudo-Goldstone boson associated with the spontaneous breaking of an approximate symmetry. Thus, we are led to choose the following potential for a complex scalar field $\chi$,
\begin{align}
V\left(\chi^\dagger,\chi\right) = \lambda\left(|\chi|^2 - v^2\right)^2 - M^2\left(\chi^\dagger\chi^\dagger + \chi\,\chi\right) + \mathbb{C},\label{potchi}
\end{align}
where $\lambda$, $v$, $M$ and $\mathbb{C}$ are constants of mass dimension $0$, $1$, $1$ and $4$, respectively. The first term in \eqref{potchi} spontaneously breaks a global $U(1)$ symmetry while the second one is a soft explicit breaking \footnote{In principle one should also consider the lower-dimensional symmetry breaking operator $\Upsilon(\chi^\dagger + \chi)$ as well. However, if we impose a $\mathbb{Z}_2$ symmetry such that $\chi \to -\chi$ leaves the action invariant,  $\Upsilon = 0$ naturally. In this work we are choosing this latter option.}. Denoting as $\widehat{\psi}$ the vacuum expectation value (VEV) of any field $\psi$, the extrema of the potential, defined through $V_{\chi^\dagger}\big|_{\left(\chi^\dagger = \widehat{\chi}^\dagger,\,\chi = \widehat{\chi}\right)} = V_{\chi}\big|_{\left(\chi^\dagger = \widehat{\chi}^\dagger,\,\chi = \widehat{\chi}\right)} = 0$, are such that $|\widehat{\chi}|^2 = v^2 \pm \frac{M^2}{\lambda}$. We will parametrize the complex scalar $\chi$ in the polar form so the (broken) symmetry is manifest, meaning
\begin{align}
\chi \equiv \frac{1}{\sqrt{2}}\left(\widetilde{R} + \sigma\right)e^{i\theta},\label{polarchi}
\end{align}
where $\widetilde{R}$ is a constant of mass dimension $1$. In the effective theory, after integrating the radial field $\sigma$, we want to recover a chaotic (concave) potential for the ``inflaton'' field $\theta$. The minimum is then taken to be
\begin{align}
|\widehat{\chi}|^2 = v^2 + \frac{M^2}{\lambda} = \frac{1}{2}\,(\widetilde{R} + \widehat{\sigma})^2 \equiv \frac{1}{2}\,R^2, \quad \widehat{\theta} = 0.\label{min}
\end{align}
Now we fix $\mathbb{C}$ by demanding a vanishing ``cosmological constant'' at the minimum 
\begin{align}
V\left(\widehat{\chi}^\dagger,\widehat{\chi}\right) = -M^2\left(2\,v^2 + \frac{M^2}{\lambda}\right) + \mathbb{C} = 0.
\end{align}
The potential $V\left(\chi^\dagger,\chi\right)$ can then be written as
\begin{align}\label{potpol}
V(\sigma,\theta) = \mu^4\left\{\left(1 - \left(\frac{\widetilde{R} + \sigma}{\sqrt{2}\,v}\right)^2\right)^2 - \beta\left(\frac{\widetilde{R} + \sigma}{\sqrt{2}\,v}\right)^2\cos(2\theta) + \beta\left(1 + \frac{\beta}{4}\right)\right\},
\end{align}
where
\begin{align}
\mu^4 \equiv \lambda\,v^4 \quad \text{and} \quad \beta \equiv \frac{2\,M^2}{\lambda\,v^2}.
\end{align}
Note that this potential is ``non-separable'', meaning $V(\sigma,\theta) \neq V(\sigma) + V(\theta)$. Since $\mu^4$ is an overall constant that is fixed by the amplitude of the $2$-point function of the inflaton fluctuation, $\beta$ and $v$ are the only parameters that determine the dynamics of the theory. It is easy to see that in the limit $\beta \to 0$, the ``masses'' of the radial and angular fields (evaluated at the minimum of the potential) are $4\,\lambda\,v^2$ and $0$, respectively. At $\mathcal{O}(\beta)$ we find that they are given by $4\,\lambda\,v^2 - 2M^2$ and $4M^2$. From now on we will pick coordinates such that, without loss of generality, 
\begin{align}
\widehat{\sigma} = 0 \to \widetilde{R} = R = \sqrt{2\,v^2 + \frac{2M^2}{\lambda}} = \sqrt{2 + \beta}\,v.\label{sighatzero}
\end{align}
Finally we can rewrite \eqref{potpol} as 
\begin{align}
V(\sigma,\theta) = \mu^4\left\{\left(1 - \left(\sqrt{1 + \frac{\beta}{2}} + \frac{\sigma}{\sqrt{2}\,v}\right)^2\right)^2 - \beta\left(\sqrt{1 + \frac{\beta}{2}} + \frac{\sigma}{\sqrt{2}\,v}\right)^2\cos(2\theta) + \beta\left(1 + \frac{\beta}{4}\right)\right\}.\label{potpol2}
\end{align}
In \figurename{\ref{fig:defpot}} we plot the potential $V(\sigma,\theta)$ for a suitable choice of couplings. We see that it can be thought of as a ``deformed'' Mexican Hat.

\newpage

\noindent The canonical Lagrangian for the $\chi$ field is given by
\begin{align}
\mathscr{L} = -\partial_\mu\chi^\dagger\partial^\mu\chi - V\left(\chi^\dagger,\chi\right),
\end{align}
which, when written in the polar coordinates \eqref{polarchi}, takes the following form \footnote{Let us note that the original QSF model \cite{Chen:2009we,Chen:2009zp} is indeed determined by a two-field system with a Lagrangian seemingly identical to the one given in \eqref{lagran} but with $V(\sigma,\theta) = V(\sigma) + V_\text{sr}(\theta)$, i.e., the potential is assumed to be ``separable''. Moreover, $V(\sigma)$ is a potential that traps the ``isocurvaton'' at some $\sigma = \widehat{\sigma}$ but remains otherwise arbitrary while $V_\text{sr}(\theta)$ is an unspecified slow-roll potential. Our model instead, has a very specific non-separable potential given by \eqref{potpol2}. The motivation behind the original QSF model was the fact that when the inflaton trajectory moves along an arc, the action can be conveniently written in terms of polar coordinates of a circle with radius $\widetilde{R}$.}
\begin{align}
\mathscr{L} = - \frac{1}{2}\tensor{g}{^\mu^\nu}\partial_\mu\sigma\partial_\nu\sigma - \frac{1}{2}(R + \sigma)^2\tensor{g}{^\mu^\nu}\partial_\mu\theta\partial_\nu\theta - V(\sigma,\theta).\label{lagran}
\end{align}
Defining $\phi^a(t) = (\sigma(t),\,\theta(t))^T$ and $\tensor{\gamma}{_a_b}(\phi) = \text{diag}\left(1,\,(R + \sigma)^2\right)$ we may cast this class of models in the geometric language of multifield inflation. The non-vanishing Levi-Civita connection components are then given by $\tensor{\Gamma}{^\sigma_\theta_\theta} = -(R + \sigma)$ and $\tensor{\Gamma}{^\theta_\theta_\sigma} = (R + \sigma)^{-1}$ (since this is a polar coordinatization of a plane, $\tensor{\mathbb{R}}{^a_b_c_d} = 0$ trivially). Consequently the field equations \eqref{covfieldeq} become
\begin{align}
\ddot{\sigma} + 3H\dot{\sigma} - (R + \sigma)\,\dot{\theta}^2 + V_\sigma &= 0,\label{eqsigma}\\
(R + \sigma)^2\,\ddot{\theta} + 2(R + \sigma)\,\dot{\sigma}\,\dot{\theta} + 3H(R + \sigma)^2\,\dot{\theta} + V_\theta &= 0\label{eqtheta}.
\end{align}
From \eqref{lagran} and \eqref{potpol2} we see that when \textit{naively} \footnote{This procedure is rather incomplete, as care must be taken of the remnant equation of motion for $\sigma$, which now becomes a constraint equation (see appendix \ref{eftofback}).} setting $\sigma$ to its VEV, $\widehat{\sigma}$ = 0, we are left with an effective NI theory \cite{Freese:1990rb,Adams:1992bn} for the canonically normalized field $\xi \equiv R\,\theta$, whose Lagrangian is given by 
\begin{align}
\mathscr{L}_{\text{eff}\,\xi} = -\frac{1}{2}\partial_\mu\xi\,\partial^\mu\xi - \widetilde{V}\left(1 - \cos\left(\frac{\xi}{f}\right)\right),\label{natinf}
\end{align}
where $\widetilde{V} \equiv m_\xi^2\,f^2$, $m_\xi \equiv 2M$ and $f \equiv \frac{R}{2}$ \footnote{\label{boundsni}NI can be succesfully fit to data. In particular for $N_\text{e-folds} > 50$ and $n_s \approx 0.96$ one finds that $f \gtrsim 10\,M_\text{Pl} \approx 2.43\times 10^{19}\,\text{GeV}$ and $\widetilde{V} \gtrsim (10^{-2}\,M_\text{Pl})^4 \approx (2.43 \times 10^{16}\,\text{GeV})^4$. Saturating these bounds implies that $m_\xi \approx 2.43 \times 10^{13}\,\text{GeV}$ and $H \approx 1.4 \times 10^{14}\,\text{GeV}$ during the slow-roll regime, so indeed $m_\xi^2 \ll H^2$, which is a requirement of the slow-roll approximation.}. In appendix \ref{eftofback} we formally show that this is indeed the single-field effective theory to a very good approximation. 

\newpage    

\subsection{EFT for the slowly-rolling field $\theta$.}
Consider the set of equations \eqref{fried}, \eqref{eqsigma} and \eqref{eqtheta}. Let us study the regime in which time derivatives of $\sigma$ can be neglected. This would naively imply that the background trajectory is a circle in field space. We then impose that
\begin{align}
\{\ddot{\sigma},\,3H\dot{\sigma}\} \ll \{(R + \sigma)\,\dot{\theta}^2,\,V_\sigma\}\quad\text{and}\quad 2(R + \sigma)\,\dot{\sigma}\,\dot{\theta} \ll \{(R + \sigma)^2\,\ddot{\theta},\,3H(R + \sigma)^2\,\dot{\theta},\,V_\theta\}.\label{ineqs}
\end{align}
It has been argued \cite{Achucarro:2015rfa} that the kinetic coupling $\mathscr{L} \ni -\frac{1}{2}(R + \sigma)^2(\partial_\mu\theta)^2$ manifests itself through the fact that the radial field will have a minimum at $\bar{\sigma} \neq \widehat{\sigma}$ where $\widehat{\sigma}$ is a solution to \eqref{min} (and $\widehat{\sigma} = 0$ is our ``good choice of coordinates''). The inequalities in \eqref{ineqs} imply that \footnote{Recall that the single field description is possible provided the kinetic energy is dominated by the angular field, or more specifically that $\dot{\sigma}^2 + (R + \sigma)^2\,\dot{\theta}^2 = \left(\left(\frac{d\sigma}{d\theta}\right)^2 + (R + \sigma)^2\right)\dot{\theta}^2 \sim (R + \sigma)^2\,\dot{\theta}^2$, which is indeed equivalent to demand $\frac{1}{(R + \sigma)}\frac{d\sigma}{d\theta} \ll 1$.}
\begin{align}
\frac{1}{(R + \bar{\sigma})}\frac{d\bar{\sigma}}{d\theta} \ll 1.\label{conddsdt}
\end{align}
During the slow-roll regime, meaning $(R + \sigma)^2\,\ddot{\theta} \ll \{3H(R + \sigma)^2\,\dot{\theta},\,V_\theta\}$ as usual, the independent equations of motion become
\begin{align}
3H(R + \sigma)^2\,\dot{\theta} + V_\theta &= 0,\label{srapprox1}\\
(R + \sigma)\,\dot{\theta}^2 &= V_\sigma,\label{srapprox2}\\
3M_\text{Pl}^2H^2 &= V.\label{srapprox3} 
\end{align}
If $\sigma = \text{constant}$, \eqref{srapprox1} and \eqref{srapprox3} are the well-known equations that govern the slow-roll regime of a genuine single-field theory, whereas \eqref{srapprox2} can be thought of as a remnant ``constraint'', after ignoring the isocurvature field dynamics, that enforces ``centripetal equilibrium'' during an almost constant angular speed turn in field space. Using the set of equations \eqref{srapprox1}, \eqref{srapprox2} and \eqref{srapprox3} it is easy to show that the algebraic relation,
\begin{align}
(R + \sigma)^3\,V_\sigma\,V = \frac{M_\text{Pl}^2}{3}\,V_\theta^2,\label{consisrel}
\end{align}
is a consistency requirement that should hold during the slow-roll evolution. We will define $\bar{\sigma} \neq 0$ as the time-dependent ``solution'' to this equation. For the potential given by \eqref{potpol2} we have that
\begin{align}
V_\theta = 2M^2(R + \sigma)^2\sin(2\theta) \quad \text{and} \quad V_\sigma = (R + \sigma)\left[\lambda\,\{(R + \sigma)^2 - 2\,v^2\} - 2M^2\cos(2\theta)\right].\label{vsigvthet}
\end{align}
It will be useful to state that, without making any assumptions about the ``displaced'' value of $\sigma$, \eqref{vsigvthet} and \eqref{sighatzero} imply that \eqref{srapprox2} becomes 
\begin{align}
\dot{\theta}^2 &= \lambda\left[\beta\left(1 - \cos(2\theta)\right)v^2 + 2\sqrt{2 + \beta}\,v\,\sigma + \sigma^2\right]\label{srapprox22}\\
&\approx \lambda\left[2\sqrt{2}\,v\,\sigma + \sigma^2 + \beta\left(\left(1 -\cos(2\theta)\right)v^2 + \frac{\sqrt{2}}{2}\,v\,\sigma\right)\right]\quad\text{to}\quad\mathcal{O}(\beta).\nonumber
\end{align}
Let us consider now two limiting cases for the displaced value of $\sigma$.

\subsubsection{Small radial displacement.}
If we assume that $\bar{\sigma}(\theta) = \sigma_1(\theta)$, where $\sigma_1$ is a ``small'' departure from the naive VEV, the solution to \eqref{consisrel} after linearizing with respect to $\sigma$ is found to be given by \begin{align}
\sigma_1(\theta) &\approx \frac{\beta}{6\sqrt{2}}\frac{\left[M_\text{Pl}^2 - 3\,v^2 + \left(M_\text{Pl}^2 + 3\,v^2\right)\cos(2\theta)\right]}{v}\quad\text{to}\quad\mathcal{O}\left(\beta\right).\label{sig1om2}
\end{align}
In principle we can plug this solution back in the potential and find a canonical variable so that we have a single-field effective potential. However, in situations in which the solution $\bar{\sigma}(\theta)$ is a complicated function of $\theta$, it may be too difficult to follow this procedure, the main reason being that we need to find a canonical variable $\phi$ such that $(R + \bar{\sigma})\,\dot{\theta} = \dot{\phi}$. However, the system can still be solved ``semi-analytically'' as was argued in \cite{Achucarro:2015rfa}.

\subsubsection{Big radial displacement.}
As we will see in section \eqref{infpert}, if we consider the perturbations around the background model in a regime where the dynamics of the fluctuation $\delta\sigma$ is negligible in comparison with its effective mass $M_\text{eff}$, the so-called $M_\text{eff}^2 \gg H^2$ regime, the EFT for the inflaton perturbation $\delta\theta$ develops a non-trivial speed of sound $c_s$ given by
\begin{align}\label{speedandmass}
c_s^{-2} = 1 + 4\frac{\dot{\theta}_0^2}{M_\text{eff}^2}\quad\text{where}\quad M_\text{eff}^2 \equiv V_{\sigma\sigma}(\sigma_0,\theta_0) - \dot{\theta}_0^2,
\end{align}
and $\psi_0$ denotes the background value of any field $\psi$ \cite{Tolley:2009fg,Cremonini:2010ua,Achucarro:2010da}. When the potential $V(\sigma,\theta)$ is given by \eqref{potpol} we find that
\begin{align}
V_{\sigma\sigma}(\sigma_0,\theta_0) &= 3\,\lambda\,(R + \sigma_0)^2 - 2(\lambda\,v^2 + M^2\cos(2\theta_0)).\label{vsigsig}
\end{align}
Using \eqref{speedandmass}, \eqref{vsigsig}, \eqref{sighatzero} and \eqref{srapprox22} the effective mass is given by 
\begin{align}
M_\text{eff}^2 &= 2\,\lambda\left[(2 + \beta)\,v^2 + 2\sqrt{2 + \beta}\,v\,\sigma + \sigma^2\right]\label{Meff2}\\
&\approx \lambda\left[4\,v^2 + 4\sqrt{2}\,v\,\sigma + 2\,\sigma^2 + \beta\left(2\,v^2 + \sqrt{2}\,v\,\sigma\right)\right]\quad\text{to}\quad\mathcal{O}(\beta).\nonumber
\end{align}
Looking at \eqref{speedandmass} we see that $c_s^2 \ll 1 \iff 4\frac{\dot{\theta}^2}{M_\text{eff}^2} \gg 1$. This condition, using \eqref{srapprox22} and \eqref{Meff2}, is equivalent to 
\begin{align}
c_s^2 \ll 1 \iff \sigma^2 + 2\sqrt{2 + \beta}\,v\,\sigma + \left[\beta(1 - 2\cos(2\theta)) - 2\right] v^2 \gg 0,
\end{align}
which is satisfied whenever
\begin{align}
\sigma \gg \left[\sqrt{4 + 2\beta\cos(2\theta)} - \sqrt{2 + \beta}\right]v \approx \sqrt{2}\left(\sqrt{2} - 1\right)v + \frac{\beta}{2}\left(\cos(2\theta) - \frac{\sqrt{2}}{2}\right)v\quad\text{to}\quad\mathcal{O}(\beta).
\end{align}
Neglecting $\mathcal{O}(\beta)$ terms we see that when the radial field is considerably displaced from its trivial minimum, i.e., $\sigma \gg \sqrt{2}\left(\sqrt{2} - 1\right)v \approx 0.585\,v > \widehat{\sigma} = 0$, it is possible, ``dynamically'', to get $c_s^2 \ll 1$. This fact has been previously understood and emphasized \cite{Achucarro:2012yr}. Though interesting, we will not consider this big radial field displacement scenario any further. Additional developments along these lines can be found in \cite{Achucarro:2015rfa}.

\subsection{Semi-analytical approach.}\label{seman}
There is a semi-analytical way of dealing with the system of equations \eqref{srapprox1}-\eqref{srapprox3} \cite{Achucarro:2015rfa}. Recalling the usual definitions of slow-roll parameters $\epsilon \equiv -\frac{\dot{H}}{H^2}$ and $\eta \equiv \frac{\dot{\epsilon}}{\epsilon\,H}$, and defining
\begin{align}
\delta &\equiv \frac{\dot{\bar{\sigma}}}{(R + \bar{\sigma})H} = \frac{1}{(R + \bar{\sigma})}\left(\frac{d\,\bar{\sigma}}{d\,\theta}\right)\left(\frac{\dot{\theta}}{H}\right) \approx -\frac{M_\text{Pl}}{(R + \bar{\sigma})^2}\left(\frac{d\,\bar{\sigma}}{d\,\theta}\right)\sqrt{2\,\epsilon}\label{delta},
\end{align}
it is straightforward to show that 
\begin{align}
\epsilon &\approx \frac{M_\text{Pl}^2}{2(R + \bar{\sigma})^2}\left(\frac{V_\theta}{V}\right)^2,\label{eps}\\
\eta &\approx -\frac{2M_\text{Pl}^2}{(R + \bar{\sigma})^2}\left(\frac{V_{\theta\theta}}{V}\right) - \frac{2M_\text{Pl}^2}{(R + \bar{\sigma})^2}\left(\frac{d\,\bar{\sigma}}{d\,\theta}\right)\left(\frac{V_{\theta\sigma}}{V}\right) + 4\,\epsilon - 2\,\delta.\label{eta}
\end{align} 
Finally, recalling that $dN \equiv -Hdt$, we get that the number of e-folds before the end of inflation is given by
\begin{align}
N = \frac{1}{M_\text{Pl}^2}\int(R + \bar{\sigma})^2\left(\frac{V}{V_\theta}\right)d\theta,
\end{align}
stressing again that $\bar{\sigma}$ is defined as the solution to \eqref{consisrel}. The deviations from NI are due to the implicit time dependence of $\bar{\sigma} = \bar{\sigma}(\theta(t))$.  
We see that even if the reduced equations of motion demand $\delta \ll 1$, $\delta$ may still be $\mathcal{O}(\epsilon,\eta)$. Thus, even if we can neglect the derivatives of $\sigma$ at the level of the equations of motion, they may still play an important role in determining the observables of the model. Using \eqref{delta}, \eqref{eps} and \eqref{eta} with $\bar{\sigma}(\theta) = \sigma_1(\theta)$ as given by \eqref{sig1om2} and the potential given by \eqref{potpol2}, we find that
\begin{equation}\resizebox{\textwidth}{!}{
$\epsilon \equiv -\frac{\dot{H}}{H^2} \approx \frac{M_\text{Pl}^2\cot^2(\theta)}{v^2}\bigg\{1 - \frac{\beta}{72}\frac{\left[3M_\text{Pl}^4 - 3M_\text{Pl}^2\,v^2 + 18\,v^4 + 2(2M_\text{Pl}^4 - 9\,v^4)\cos(2\theta) + M_\text{Pl}^2(M_\text{Pl}^2 + 3\,v^2)\cos(4\theta)\right]\csc^2(\theta)}{v^4}\bigg\}$,}
\end{equation}
\begin{equation}\resizebox{\textwidth}{!}{
$\eta \equiv \frac{\dot{\epsilon}}{\epsilon\,H} \approx \frac{2M_\text{Pl}^2\csc^2(\theta)}{v^2}\bigg\{1 - \frac{\beta}{18}\frac{\left[(M_\text{Pl}^2 + 3\,v^2)(6M_\text{Pl}^2 - 3\,v^2 + M_\text{Pl}^2\cos(2\theta))\sin^2(\theta) - 8M_\text{Pl}^4 - 12M_\text{Pl}^2\,v^2 + 9\,v^4 + 3M_\text{Pl}^4\csc^2(\theta)\right]}{v^4}\bigg\}$.}
\end{equation}
We see that to $\mathcal{O}(\beta^0)$, $\{\epsilon, \eta\} \sim \frac{M_\text{Pl}^2}{v^2}$ which implies that in order to have $\{\epsilon, \eta\} \ll 1$ we need $v^2 \gg M_\text{Pl}^2$, as is usually the case for NI.\\ 
To compare the predictions of the model with data we recall the well-known formulae of the scalar spectral index $n_s$ and the tensor-to-scalar ratio $r$, i.e. $n_s - 1 = -2\,\epsilon - \eta - s$, where $s \equiv \frac{\dot{c}_s}{c_s\,H}$, and $r = 16\,\epsilon\,c_s$ (see \cite{Baumann:2009ds} for standard definitions). In our model $\{\dot{\sigma} \approx 0 \to \dot{\theta} \approx \text{constant}\} \to \{c_s \approx \text{constant} \to s \approx 0\}$ to $\mathcal{O}(\beta^0)$. We pick parameter values $\lambda$, $v$ and $M$ so they are compatible with the set of relations
\begin{align}\label{setofcond}
&\widetilde{V} \approx \frac{3\,\pi^2}{2}\,r\,\Delta_\mathcal{R}^2\,M_\text{Pl}^4,\quad v \approx \sqrt{\frac{16\,c_s}{r}}M_\text{Pl},\quad M \approx \sqrt{\frac{\widetilde{V}}{2\,v^2}},\\
&\quad H \approx \sqrt{\frac{\widetilde{V}}{3}}\,M_\text{Pl}^{-1},\quad \lambda \equiv \frac{\widetilde{\alpha}\,(1 + 3\,c_s^2)H^2}{16\,v^2},\quad \dot{\theta} \approx \sqrt{\frac{4\,\lambda\,v^2\,\left(1 - c_s^2\right)}{1 + 3\,c_s^2}},\nonumber
\end{align}
where $\widetilde{V}^{1/4}$ is the energy scale of inflation, $\Delta_\mathcal{R}^2 = 2.14 \times 10^{-9}$ is the (measured) dimensionless power spectrum of the inflaton perturbation, $\widetilde{\alpha} \equiv \frac{\widetilde{M}_\text{eff}^2}{H^2}$ is the ratio between the ``Hamiltonian effective mass squared'' $\widetilde{M}_\text{eff}^2 \equiv M_\text{eff}^2\,c_s^{-2}$ and $H^2$ and we are neglecting $\mathcal{O}(\beta)$ terms \footnote{In subsection \ref{eftbreaks} and appendix \ref{efftheory} the introduction of the more ``physical'' effective mass $\widetilde{M}_\text{eff}$ is justified. $\widetilde{M}_\text{eff}$ has been also referred to as the ``entropy mass'' \cite{Gordon:2000hv} and it really corresponds to the mass of a particle belonging to the spectrum of the theory, which is not the case for $M_\text{eff}$.}. Indeed, $\beta \equiv \frac{2\,M^2}{\lambda\,v^2} \approx \frac{3}{1 + 3\,c_s^2}\frac{r}{\widetilde{\alpha}\,c_s} \approx \frac{48}{1 + 3\,c_s^2}\,\frac{\epsilon}{\widetilde{\alpha}}$ within the above approximations, so $\beta$ is always a very small number due to slow-roll. Then, we saturate the current constraint $r < 0.07$ \cite{Array:2015xqh} in \eqref{setofcond} to build up \figurename{\ref{fig:rvsns}} and Table \ref{Table1} below. Note that when $\widetilde{\alpha} \sim \mathcal{O}(1)$ the EFT for a single-field theory is not really justified since $\widetilde{M}_\text{eff}^2 \sim H^2$. Nevertheless, it is illuminating to ``extrapolate'' our results since, in particular, the value of $\frac{\dot{\theta}}{H}$ is quite important for the theory of fluctuations exactly in this limit, as we will see in the next section.
\begin{figure}[h!]
      \includegraphics[width=\textwidth]{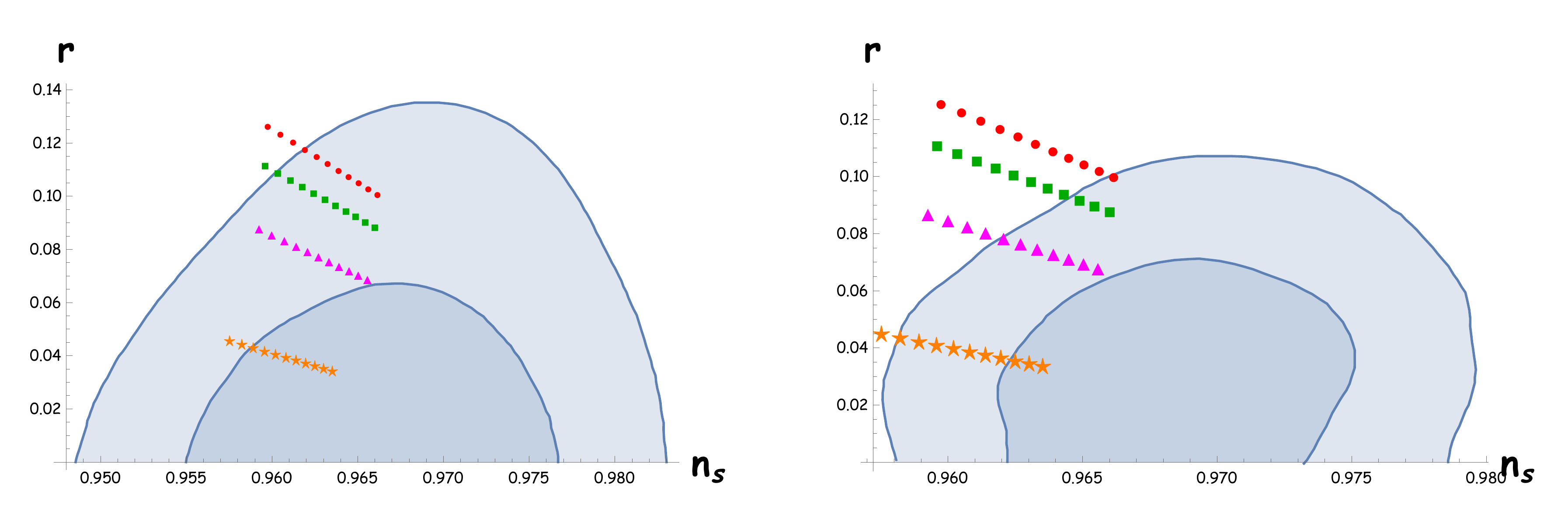}
      \caption{The ($n_s,r$) plane for the ``Natural'' potential, when the mass of the heavy field is given by $\widetilde{M}_\text{eff}^2 = 100\,H^2$. The blue regions are the 1-$\sigma$ and 2-$\sigma$ allowed regions from \textit{Left:} Planck 2015 (Planck TT+lowP) \cite{Ade:2015xua} and \textit{Right:} Planck/Bicep (Planck TT+lowP+BKP+lensing+ext) \cite{Array:2015xqh}.\\
      We plot the predictions for $N = [50,60]$ when $c_s = 0.999$ ($\color{red}{\bullet}$), $c_s = 0.9$ ($\color{ForestGreen}{\blacksquare}$), $c_s = 0.75$ ($\color{magenta}{\blacktriangle}$) and $c_s = 0.47$ ($\color{orange}{\bigstar}$).}
      \label{fig:rvsns}
\end{figure}
\begin{table}[h!]
\resizebox{\textwidth}{!}{
\begin{tabular}{|c|c|c|c|c|c|c|c|c|c|}
\cline{2-10}
\multicolumn{1}{c|}{} & $\bf{\lambda}_{100}$ & $\bf{\dot{\theta}}_{100}\,(\text{GeV})$ & $(\bf{\dot{\theta}}/\bf{H})_{100}$ & $\bf{\lambda}_{10}$ & $\bf{\dot{\theta}}_{10}\,(\text{GeV})$ & $(\bf{\dot{\theta}}/\bf{H})_{10}$ & $\bf{\lambda}_{1}$ & $\bf{\dot{\theta}}_{1}\,(\text{GeV})$ & $(\bf{\dot{\theta}}/\bf{H})_{1}$ \\
\hline
$\bf{\color{red}{c_s = 0.999}}$ & $8.081\times10^{-11}$ & $1.477\times10^{13}$ & 0.224 & $8.081\times10^{-12}$ & $4.671\times10^{12}$ & 0.071 & $8.081\times10^{-13}$ & $1.477\times10^{12}$ & 0.022 \\
\hline
$\bf{\color{ForestGreen}{c_s = 0.9}}$ & $7.704\times10^{-11}$ & $1.440\times10^{14}$ & 2.179 & $7.704\times10^{-12}$ & $4.553\times10^{13}$ & 0.689 & $7.704\times10^{-13}$ & $1.440\times10^{13}$ & 0.218 \\
\hline
$\bf{\color{magenta}{c_s = 0.75}}$ & $7.243\times10^{-11}$ & $2.185\times10^{14}$ & 3.307 & $7.243\times10^{-12}$ & $6.910\times10^{13}$ & 1.046 & $7.243\times10^{-13}$ & $2.185\times10^{13}$ & 0.331 \\
\hline
$\bf{\color{orange}{c_s = 0.47}}$ & $7.151\times10^{-11}$ & $2.916\times10^{14}$ & 4.413 & $7.151\times10^{-12}$ & $9.221\times10^{13}$ & 1.396 & $7.151\times10^{-13}$ & $2.916\times10^{13}$ & 0.441 \\
\hline
\end{tabular}}
\caption{$\lambda_{\widetilde{\alpha}}$, $\dot{\theta}_{\widetilde{\alpha}}$ and $\left(\frac{\dot{\theta}}{H}\right)_{\widetilde{\alpha}}$, where $X_{\widetilde{\alpha}} \equiv X(\widetilde{\alpha})$, with $\widetilde{\alpha} \equiv \frac{\widetilde{M}_\text{eff}^2}{H^2} = \{100,10,1\}$ for different values of $c_s$.}
\label{Table1}
\end{table}

\noindent Looking at \figurename{\ref{fig:rvsns}} we see that for $N = 60$, this model is alive and well, meaning the current constraint $r < 0.07$ is satisfied \cite{Array:2015xqh}, when $c_s = 0.75$. The only parameter of the model which depends upon $\widetilde{\alpha}$ is $\lambda$, which only influences the slow-roll parameters (therefore the predictions for the observables) at a negligible order way beyond the current experimental sensitivity. In other words, taking $\widetilde{\alpha} = \{100,10,1\}$ gives the same predictions depicted in \figurename{\ref{fig:rvsns}}. However it is interesting to note from Table \ref{Table1} that for fixed $c_s$, as $\widetilde{\alpha}$ decreases, $\frac{\dot{\theta}}{H}$ decreases too, since $\frac{\dot{\theta}}{H} \approx \frac{1}{2}\,\widetilde{\alpha}^{1/2}\left(1 - c_s^2\right)^{1/2}$ according to \eqref{setofcond}. This feature is doubly reassuring:

\newpage

\noindent (1) It is consistent with the fact that according to the EFT analysis it has been understood that $\dot{\theta}^2 \ll H^2$ is \textit{not} a restriction for the EFT to be valid as some authors initially argued in the literature \footnote{In \cite{Cespedes:2012hu} the ``adiabaticity'' condition $|\ddot{\theta}| \ll M_\text{eff}\,|\dot{\theta}|$ has been identified as a requirement for the heavy field to not become excited during the turn.}.\\
(2) When the heavy field is not super heavy, like in the QSF scenario, $\frac{\dot{\theta}}{H}$ plays the role of a time-dependent coupling between the adiabatic and isocurvature perturbations, so  $\dot{\theta}^2 \ll H^2$ is a standard perturbative condition one needs to impose to do perturbative physics. Even if the limit $\widetilde{\alpha}\to1$ is ill-defined from the single-field EFT point of view, we believe this extrapolation sheds some light on the perturbative limitations that the theory of fluctuations has in the two-field regime (see \ref{eftbreaks} below). Let us now study the theory of fluctuations.

\section{Inflationary Perturbations in the GNI Model.}\label{infpert}
In this section we will study the theory of fluctuations of the GNI model in order to calculate the non-Gaussianities that arise due to the presence of the isocurvature mode. We will address the regimes $M_\text{eff}^2 \gg H^2$ and $M_\text{eff}^2 \sim H^2$ separately, as the physics is quite different.\\   
To study the inflationary perturbations defined as $\delta\phi^a(t,\mathbf{x}) \equiv \phi^a(t,\mathbf{x}) - \phi^a_0(t)$ it is useful to consider vectors tangent and normal to the trajectory $\phi^a_0(t)$ given by
\begin{align}\label{vectors}
T^a \equiv \frac{\dot{\phi}^a_0}{\dot{\phi}_0}, \quad N^a \equiv -\frac{D_t\,T^a}{|D_t\,T|}.
\end{align}
The fluctuations along the direction $T^a$ define the curvature perturbations as $\mathcal{R} \equiv -\frac{H}{\dot{\phi}_0}\,T_a\,\delta\phi^a$ (see footnote \ref{Rmeaning} and appendix \ref{eftofinf} below) whereas the fluctuations along $N^a$ correspond to the isocurvature perturbations \cite{Gordon:2000hv,GrootNibbelink:2001qt}. The introduction of $T^a$ and $N^a$ allows us to define $\Omega$, the angular velocity with which the inflationary trajectory bends, via
\begin{align}\label{Omega}
D_t\,T^a \equiv -\Omega\,N^a.
\end{align}
Comparing \eqref{vectors} with \eqref{Omega} we see that $\Omega = |D_t\,T|$ is positive definite by construction. It is clear that in the two-field case $\{T^a,N^a\}$ is an orthonormal basis that spans the vector space, implying that $V_a = V_\phi\,T_a + V_NN_a$, where $V_\phi \equiv T^aV_a$ and $V_N \equiv N^aV_a$. The equation resulting from projecting \eqref{covfieldeq} along $T^a$ is 
\begin{align}
\ddot{\phi}_0 + 3H\dot{\phi}_0 + V_\phi = 0,\label{projalongt}
\end{align}
resembling the equation of motion of a single scalar field in a FLRW spacetime. On the other hand, the equation obtained from projecting \eqref{covfieldeq} along $N^a$ is given by
\begin{align}
\Omega = \frac{V_N}{\dot{\phi}_0}.\label{defangvel}
\end{align}
Whenever the trajectory is subjected to a bend, it moves up towards the outer wall of the potential. The angular velocity $\Omega$ plays a crucial role in the dynamics of fluctuations, as it couples together curvature and isocurvature modes. From \eqref{vectors} we see that the normal vector is constructed such that $T_aN^a = 0$ and $N_aN^a = 1$. In the two-field case it can be taken as $N_a = (\text{det}\,\gamma)^{1/2}\,\tensor{\varepsilon}{_a_b}\,T^b$, where $\tensor{\varepsilon}{_a_b}$ is the two-dimensional Levi-Civita symbol with $\tensor{\varepsilon}{_1_1} = \tensor{\varepsilon}{_2_2} = 0$ and $\tensor{\varepsilon}{_1_2} = -\tensor{\varepsilon}{_2_1} = 1$. Then for our model we get that \cite{Achucarro:2015caa}
\begin{align}
T^a = \frac{(\dot{\sigma}_0,\,\dot{\theta}_0)}{\left[(R + \sigma_0)^2\,\dot{\theta}_0^2 + \dot{\sigma}_0^2\right]^{1/2}}, \quad N^a = \frac{(R + \sigma_0)(\dot{\theta}_0,\,-(R + \sigma_0)^{-2}\dot{\sigma}_0)}{\left[(R + \sigma_0)^2\,\dot{\theta}_0^2 + \dot{\sigma}_0^2\right]^{1/2}}.
\end{align}
Considering \eqref{defangvel} we see that since $V_N \equiv \Omega\,\dot{\phi}_0 = N^\sigma V_\sigma + N^\theta V_\theta$, $\Omega$ is given by
\begin{align}
\Omega = \frac{1}{\left[(R + \sigma_0)^2\,\dot{\theta}_0^2 + \dot{\sigma}_0^2\right]}&\bigg\{(R + \sigma_0)\,\dot{\theta}_0\left(-\ddot{\sigma}_0 - 3H\dot{\sigma}_0 + (R + \sigma_0)\,\dot{\theta}_0^2\right)\nonumber\\
&- (R + \sigma_0)^{-1}\dot{\sigma}_0\left(-(R + \sigma_0)^2\,\ddot{\theta}_0 - 2(R + \sigma)\,\dot{\sigma}_0\,\dot{\theta}_0 - 3H(R + \sigma_0)^2\,\dot{\theta}_0\right)\bigg\},
\end{align}
where use has been made of \eqref{eqsigma} and \eqref{eqtheta}. Thus,
\begin{align}
\text{if} \quad \sigma_0 = \text{constant}, \quad \text{meaning} \quad \dot{\sigma}_0 = 0, \quad \text{then} \quad \Omega = \dot{\theta}_0,
\end{align}
without assuming slow-roll conditions on $\theta_0$.\\
The theory of fluctuations of the polar fields is determined by the expansion \footnote{As usual, the $S^{(1)}[g_0,\phi_0,\delta\phi]$ term in this expansion vanishes due to the background equations of motion \eqref{eqsigma} and \eqref{eqtheta}.}
\begin{align}
S[g_0,\phi_0,\delta\phi] &= S^{(0)}[g_0,\phi_0] + S^{(2)}[g_0,\phi_0,\delta\phi] + S^{(3)}[g_0,\phi_0,\delta\phi] + \dots,\label{Sfluc}\\
S^{(0)}[g_0,\phi_0] &= \int d^4x\,a^3\left(\frac{1}{2}(R + \sigma_0)^2\,\dot{\theta}_0^2 + \frac{1}{2}\dot{\sigma}_0^2 - V(\sigma_0,\theta_0)\right),\\
S^{(2)}[g_0,\phi_0,\delta\phi] &= \int d^4x\,a^3\bigg(-\frac{1}{2}(R + \sigma_0)^2\tensor{g}{^\mu^\nu}\partial_\mu\delta\theta\partial_\nu\delta\theta - \frac{1}{2}V_{\theta\theta}(\sigma_0,\theta_0)(\delta\theta)^2\nonumber\\ 
&+ 2(R + \sigma_0)\,\dot{\theta}_0\delta\dot{\theta}\delta\sigma - V_{\theta\sigma}(\sigma_0,\theta_0)\delta\theta\delta\sigma - \frac{1}{2}\tensor{g}{^\mu^\nu}\partial_\mu\delta\sigma\partial_\nu\delta\sigma - \frac{1}{2}M_\text{eff}^2(\delta\sigma)^2\bigg),\label{S2}\\
S^{(3)}[g_0,\phi_0,\delta\phi] &= \int d^4x\,a^3\bigg(-(R + \sigma_0)(\tensor{g}{^\mu^\nu}\partial_\mu\delta\theta\partial_\nu\delta\theta)\delta\sigma + \dot{\theta}_0\delta\dot{\theta}(\delta\sigma)^2 - \frac{1}{6}V_{\theta\theta\theta}(\sigma_0,\theta_0)(\delta\theta)^3\nonumber\\
&- \frac{1}{2}V_{\theta\sigma\sigma}(\sigma_0,\theta_0)\delta\theta(\delta\sigma)^2 - \frac{1}{2}V_{\theta\theta\sigma}(\sigma_0,\theta_0)(\delta\theta)^2\delta\sigma - \frac{1}{6}V_{\sigma\sigma\sigma}(\sigma_0,\theta_0)(\delta\sigma)^3\bigg),\label{S3}
\end{align}
where $M_\text{eff}^2 \equiv V_{\sigma\sigma}(\sigma_0,\theta_0) - \dot{\theta}_0^2$ as in \eqref{speedandmass} and the $\dots$ in \eqref{Sfluc} stem from higher order terms in the expansion. Let us now consider the $M_\text{eff}^2 \gg H^2$ scenario.

\subsection{$M_\text{eff}^2 \gg H^2$ regime.}\label{mgghregime}
\subsubsection{Effective Theory for the adiabatic (inflaton) fluctuation.}\label{eftadiab}
In this subsection we will show how the naive expectation, that when the mass of the isocurvature mode is very heavy we can integrate it out to obtain an effective single-field description with non-trivial coefficients for non-slow-roll operators, is realized. We will match our findings with the general parametrization introduced in the so-called EFT of inflation developed by Cheung et al. \cite{Cheung:2007st}, for which the relations between coefficients of the EFT and the amplitudes of non-Gaussianities are well-known.\\
Following Gong et al. \cite{Gong:2013sma} we vary \eqref{S2} and \eqref{S3} with respect to $\delta\sigma$ to obtain 
\begin{align}
\delta\ddot{\sigma} + 3H\delta\dot{\sigma} - &\left(\frac{\nabla^2}{a^2} - M_\text{eff}^2 + 2\dot{\theta}_0\,\delta\dot{\theta} - V_{\theta\sigma\sigma}\delta\theta\right)\delta\sigma + \frac{V_{\sigma\sigma\sigma}}{2}(\delta\sigma)^2\nonumber\\
&= 2(R + \sigma_0)\,\dot{\theta}_0\,\delta\dot{\theta} - V_{\theta\sigma}\,\delta\theta + (R + \sigma_0)\left((\delta\dot{\theta})^2 - \frac{(\nabla\delta\theta)^2}{a^2}\right) - \frac{V_{\theta\theta\sigma}}{2}(\delta\theta)^2.
\end{align}
Assuming that the effective mass of $\delta\sigma$ is very large (so the term $M_\text{eff}^2\,\delta\sigma$ dominates in the above equation) and neglecting its dynamics, we can find a perturbative solution given by \footnote{Here we have neglected both time derivatives and gradients of $\delta\sigma$. In principle, one can keep the gradients to obtain an effective theory that captures the regime of non-linear dispersion relations \cite{Gwyn:2012mw,Castillo:2013sfa}, the so-called ``new physics window'' dubbed by Baumann and Green \cite{Baumann:2011su} (see appendix \ref{efftheory} to get a quick understanding of how non-linear dispersion relations generically arise when integrating out a heavy field). In \eqref{constrsol} we are also neglecting terms proportional to $M^2$ since $M^2 \ll M_\text{eff}^2$.}
\begin{align}
\delta\sigma \approx \frac{2R\,\dot{\theta}_0}{M_\text{eff}^2}\delta\dot{\theta} + \left(\frac{R}{M_\text{eff}^2} - \frac{2R^2\,\dot{\theta}_0^2}{M_\text{eff}^2}\frac{V_{\sigma\sigma\sigma}}{M_\text{eff}^4}\right)(\delta\dot{\theta})^2,\label{constrsol}
\end{align}
where we have taken $\sigma_0 = \widehat{\sigma} = 0$.\\
Plugging \eqref{constrsol} back into \eqref{S2} and \eqref{S3}, and keeping only the leading order terms in slow-roll parameters, we find the effective single field fluctuation action
\begin{align}
S^{(2)}_{\text{eff}\,\delta\theta}[g_0,\theta_0,\delta\theta] &= \int d^4x\,a^3\bigg(\frac{1}{2}R^2(\delta\dot{\theta})^2\bigg(1 + 4\frac{\dot{\theta}_0^2}{M_\text{eff}^2}\bigg) - \frac{1}{2}R^2\frac{(\nabla\delta\theta)^2}{a^2}\bigg),\\
S^{(3)}_{\text{eff}\,\delta\theta}[g_0,\theta_0,\delta\theta] &= \int d^4x\,a^3\bigg(\bigg[\frac{R^2\,\dot{\theta}_0}{M_\text{eff}^2} + \frac{R^2\,\dot{\theta}_0}{M_\text{eff}^2}\bigg(1 + 4\frac{\dot{\theta}_0^2}{M_\text{eff}^2}\bigg) - \frac{4}{3}\frac{R^3\,\dot{\theta}_0^3}{M_\text{eff}^6}V_{\sigma\sigma\sigma}\bigg](\delta\dot{\theta})^3\nonumber\\
&- \frac{2R^2\,\dot{\theta}_0}{a^2M_\text{eff}^2}\delta\dot{\theta}(\nabla\delta\theta)^2\bigg).
\end{align}
Indeed we see that if we define the speed of sound $c_s$ through \eqref{speedandmass}, the quadratic action is equivalent to that of general single-field inflation. To evaluate the observable quantities, we have to transfer this action into that of the curvature perturbation. It is well known that the curvature perturbation on the comoving slices $\mathcal{R}$ \footnote{\label{Rmeaning}Recall that $\mathcal{R}$ is the gauge invariant quantity that does not evolve on super-Hubble scales $\frac{k}{a} \ll H$ (or super-sound-horizon crossing scales $k\,c_s \ll aH$ if $c_s \neq 1$), unless non-adiabatic pressure is significant. This fact is of course crucial for relating late-time observables, such as the distribution of galaxies, to the initial conditions from inflation. See \cite{Baumann:2014nda} and appendix \ref{eftofinf} below.}  is given in terms of the field fluctuation on the flat slices along the trajectory $\delta\theta$ as 
\begin{align}
\mathcal{R} = -\frac{H}{\dot{\theta}_0}\,\delta\theta.
\end{align}
A straightforward calculation shows that
\begin{align}
S^{(2)}_{\text{eff}\,\mathcal{R}}[g_0,\theta_0,\mathcal{R}] &= M_\text{Pl}^2\int d^4x\,a^3\frac{\epsilon}{c_s^2}\left[\dot{\mathcal{R}}^2 - c_s^2\frac{(\nabla\mathcal{R})^2}{a^2}\right]\label{Reffaction2}\\
S^{(3)}_{\text{eff}\,\mathcal{R}}[g_0,\theta_0,\mathcal{R}] &= M_\text{Pl}^2\int d^4x\,a^3\bigg(-\frac{H^2\epsilon}{c_s^2}\left[\frac{c_s^2}{2}\left(\frac{1}{c_s^4} - 1\right) - c_s^2\frac{R\,V_{\sigma\sigma\sigma}}{6M_\text{eff}^2}\left(\frac{1}{c_s^2} - 1\right)^2\right]\frac{\dot{\mathcal{R}^3}}{H^3}\nonumber\\
&+ \epsilon\left(\frac{1}{c_s^2} - 1\right)\frac{\dot{\mathcal{R}}}{H}\frac{(\nabla\mathcal{R})^2}{a^2}\bigg),\label{Reffaction3}
\end{align}
where $\epsilon \equiv -\frac{\dot{H}}{H^2} = \frac{R^2\,\dot{\theta}_0^2}{2M_\text{Pl}^2H^2}$. 

\newpage

\noindent We see that $\mathcal{R}$ is indeed massless which implies that $\dot{\mathcal{R}} \approx 0$ at super-sound-horizon crossing scales, $k\,c_s \ll aH$ \cite{Weinberg:2003sw}.\\
Now let us recall that the effective action for the Goldstone boson $\pi$ of gravity in a de Sitter background reads \cite{Cheung:2007st,Creminelli:2006xe} (see appendix \ref{eftofinf} for details)
\begin{align} 
S_\pi = \int d^4x\,a^3\bigg(-M_\text{Pl}^2\dot{H}\left[\dot{\pi}^2 - \frac{(\nabla\pi)^2}{a^2}\right] + 2M_2^4\left[\dot{\pi}^2 + \dot{\pi}^3 - \dot{\pi}\frac{(\nabla\pi)^2}{a^2}\right] - \frac{4}{3}M_3^4\,\dot{\pi}^3 + \dots\bigg),\label{pigravity}
\end{align}
where $M_2(t)$ and $M_3(t)$ are (a priori) undetermined time-dependent coefficients of mass dimension 1. From \eqref{pigravity} we see that the speed of sound of $\pi$ fluctuations is given by
\begin{align}
(c_s^\pi)^{-2} = 1 - \frac{2M_2^4}{M_\text{Pl}^2\dot{H}},
\end{align}
so the Goldstone action can be rewritten at cubic order as
\begin{align}\label{Goldseffaction}
S_\pi = \int d^4x\,a^3\bigg(-\frac{M_\text{Pl}^2\dot{H}}{(c_s^\pi)^2}\left[\dot{\pi}^2 - (c_s^\pi)^2\frac{(\nabla\pi)^2}{a^2}\right] + M_\text{Pl}^2\dot{H}\left(1 - \frac{1}{(c_s^\pi)^2}\right)\left[\dot{\pi}^3 - \dot{\pi}\frac{(\nabla\pi)^2}{a^2}\right] - \frac{4}{3}M_3^4\,\dot{\pi}^3 + \dots\bigg).
\end{align}
Using the fact that $\mathcal{R} = -H\,\pi + \mathcal{O}(\pi^2)$ (see appendix \ref{eftofinf}) and identifying $c_s^\pi = c_s$, we find, comparing \eqref{Reffaction3} with \eqref{Goldseffaction} that 
\begin{align}
M_2^4 &= \frac{1}{2}\,\epsilon\,M_\text{Pl}^2H^2\left(\frac{1}{c_s^2} - 1\right),\label{M2}\\
M_3^4 &= \frac{3}{4}\,\epsilon\,M_\text{Pl}^2H^2\left(\frac{1}{c_s^2} - 1\right)^2\left[\frac{R}{6M_\text{eff}^2}V_{\sigma\sigma\sigma} - \frac{1}{2}\right].\label{M3}
\end{align} 
It can be shown that in the limit when self-interactions of the heavy field $\sigma$ are ignored while ``solving'' its own (constraint) equation of motion, the sound speed $c_s$ and the couplings $M_n^4$ are uniquely related by \cite{Achucarro:2012sm}
\begin{align}\label{coeffnaiveeft}
 M_n^4 = (-1)^n\,n!|\dot{H}|M_\text{Pl}^2\left(\frac{c_s^{-2}-1}{4}\right)^{n-1}.
\end{align}
Indeed, we see from \eqref{M2} and \eqref{M3} that if the $V_{\sigma\sigma\sigma}$ term is dropped we agree with this result. Comparing the coefficient $M_3^4 \sim M_\text{eff}^{-4}$ coming from \eqref{coeffnaiveeft} to the one calculated in \eqref{M3}, which has an additional $\sim M_\text{eff}^{-6}$ behavior, we realize that $M_3^4$ reflects the non-linear self-interaction of the heavy field during inflation as was stressed in \cite{Gong:2013sma}. This is based on the fact that the $V_{\sigma\sigma\sigma}$ term actually dominates $M_3$ even if it is naively further suppressed by one more power of $M_\text{eff}^2$\,. This fact is discussed by the end of appendix \ref{nongaussdef}. We now estimate and calculate non-Gaussianities arising in this particular limit or our model.

\subsubsection{Non-Gaussianities.}

In order to \textit{estimate} the non-Gaussianities associated with the effective action for $\mathcal{\pi}$ it is convenient to absorb the sound speed into a redefinition of the spatial coordinates $x^i\to\tilde{x}^i \equiv c_s^{-1}\,x^i$ so that ``fake'' Lorentz invariance is restored \cite{Baumann:2011su,Baumann:2014nda}. Then the effective theory Lagrangian $\widetilde{\mathscr{L}}_\pi \equiv c_s^3\,\mathscr{L}_\pi$ can be casted like
\begin{align}
\widetilde{\mathscr{L}}_\pi = -\frac{1}{2}(\widetilde{\partial}_\mu\pi_c)^2 - \frac{1}{2\,\Lambda^2}\left(\dot{\pi}_c\,\frac{(\widetilde{\nabla}\pi_c)^2}{a^2} + \mathcal{A}\,\dot{\pi}_c^3\right),
\end{align}
where $\widetilde{\partial}_\mu \equiv (\partial_t,\,c_s\,\partial_i)$, $\pi_c \equiv f_\pi^2\,\pi$ is a canonically normalized field and
\begin{align}
f_\pi^4 &\equiv 2M_\text{Pl}^2|\dot{H}|c_s,\label{fpi}\\
\Lambda^4 &\equiv \frac{c_s^4}{(1 - c_s^2)^2}\,f_\pi^4 = \frac{2M_\text{Pl}^2|\dot{H}|c_s^5}{(1 - c_s^2)^2},\label{Lambda}\\
\frac{\mathcal{A}}{c_s^2} &\equiv -1 + \frac{2}{3}\frac{M_3^4}{M_2^4}.\label{Adeff}
\end{align}
Here, $f_\pi^4$ and $\Lambda^4$ are the so-called ``symmetry breaking'' and ``strong coupling'' \footnote{It can be shown that the breakdown of perturbative unitarity of Goldstone boson scattering occurs when $\omega^4 > \frac{24\pi}{5}(1 - c_s^2)\Lambda^4 \equiv \Lambda_u^4$ \cite{Baumann:2011su,Baumann:2014cja} . $\Lambda_u$ is referred to as the ``unitarity bound''. These definitions rely on the linear dispersion relation that we have assumed throughout this work.} scales respectively. A simple ``back-of-the-envelope'' estimate for the amplitude of the non-Gaussianity can be found by comparing the non-linear (cubic) terms with the quadratic terms in the Lagrangian, around freezing time $\omega \sim H$. This is because the interaction operators have derivatives acting on the fluctuations so they effectively are shut down after freezing. Using our fake Lorentz-invariant Lagrangian we find that 
\begin{align}
f_\text{NL}^{\dot{\pi}(\nabla\pi)^2}\mathcal{R} &\equiv \frac{\widetilde{\mathscr{L}}_3^{\,\,\dot{\pi}(\nabla\pi)^2}}{\widetilde{\mathscr{L}}_2}\bigg|_{\omega \sim H} \sim \frac{1}{2\,\Lambda^2}\frac{\dot{\pi}_c\,(\widetilde{\partial}\,\pi_c)^2}{(\widetilde{\partial}\,\pi_c)^2} \sim \left(\frac{f_\pi}{\Lambda}\right)^2\mathcal{R} \sim \left(\frac{1 - c_s^2}{c_s^2}\right)\mathcal{R},\label{estfnltim}\\
f_\text{NL}^{\dot{\pi}^3}\mathcal{R} &\equiv \frac{\widetilde{\mathscr{L}}_3^{\,\,\dot{\pi}^3}}{\widetilde{\mathscr{L}}_2}\bigg|_{\omega \sim H} \sim \frac{\mathcal{A}}{2\,\Lambda^2}\frac{\dot{\pi}_c^3}{(\widetilde{\partial}\,\pi_c)^2} \sim \mathcal{A}\left(\frac{f_\pi}{\Lambda}\right)^2\mathcal{R} \sim \mathcal{A}\left(\frac{1 - c_s^2}{c_s^2}\right)\mathcal{R}\label{estfnlgrad},
\end{align}
where the exact definition of $f_\text{NL}^{\dot{\pi}^3}$ is given below in equation \eqref{fnldef} (there is of course an equivalent definition for $f_\text{NL}^{\dot{\pi}(\nabla\pi)^2}$). Then it is easy to estimate non-Gaussianities once the matching between a particular model and the EFT of inflation has been made. 

\newpage

\begin{figure}[h]
\begin{center}
      \includegraphics[width=\textwidth]{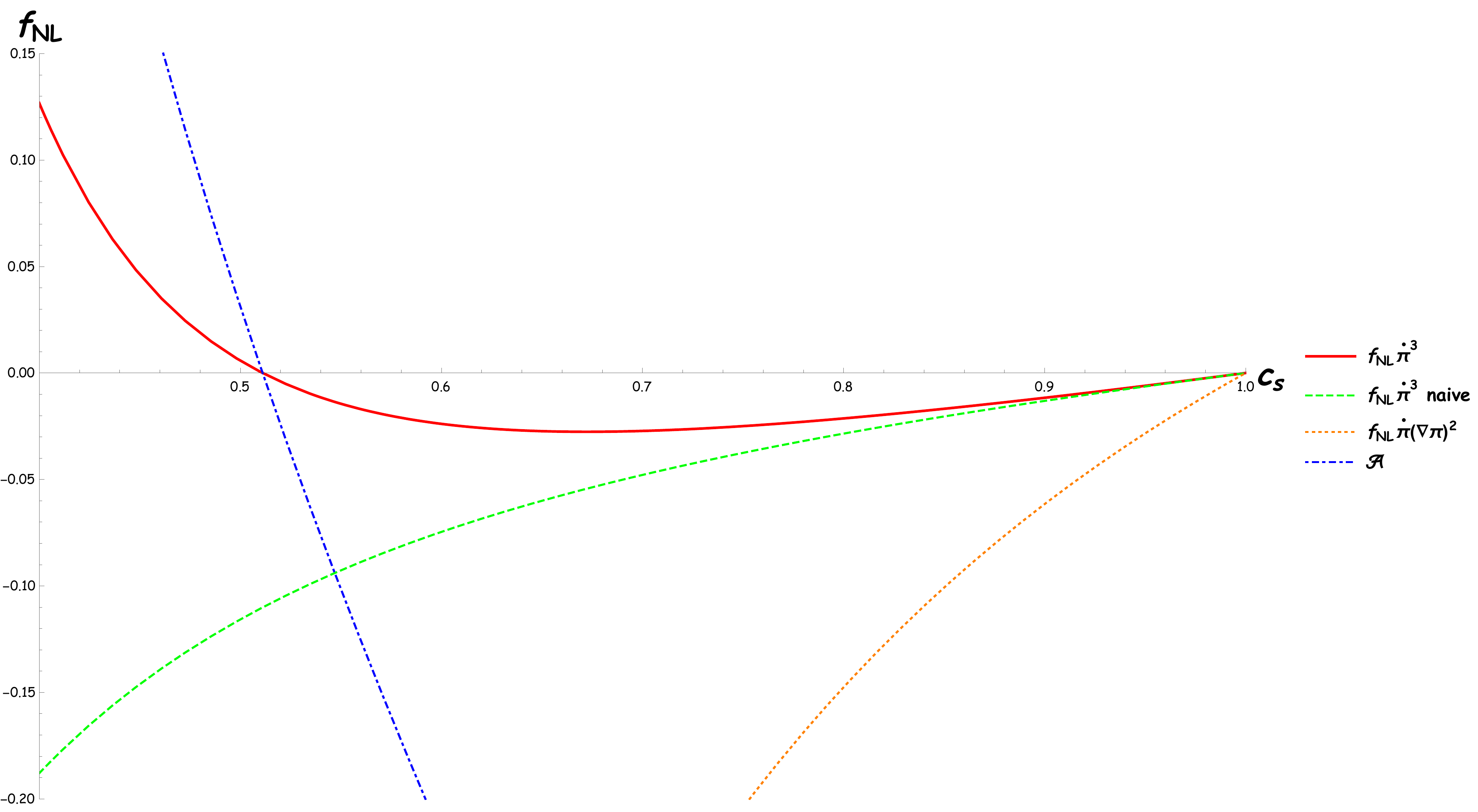}
      \caption{$f_\text{NL}^{\dot{\pi}^3}$, $f_\text{NL\,(naive)}^{\dot{\pi}^3}$, $f_\text{NL}^{\dot{\pi}(\nabla\pi)^2}$ and $\mathcal{A}$ vs. $c_s$.}
      \label{fig:fnlpivscs}
\end{center}
\end{figure}

\begin{figure}[h]
\begin{center}
      \includegraphics[width=\textwidth]{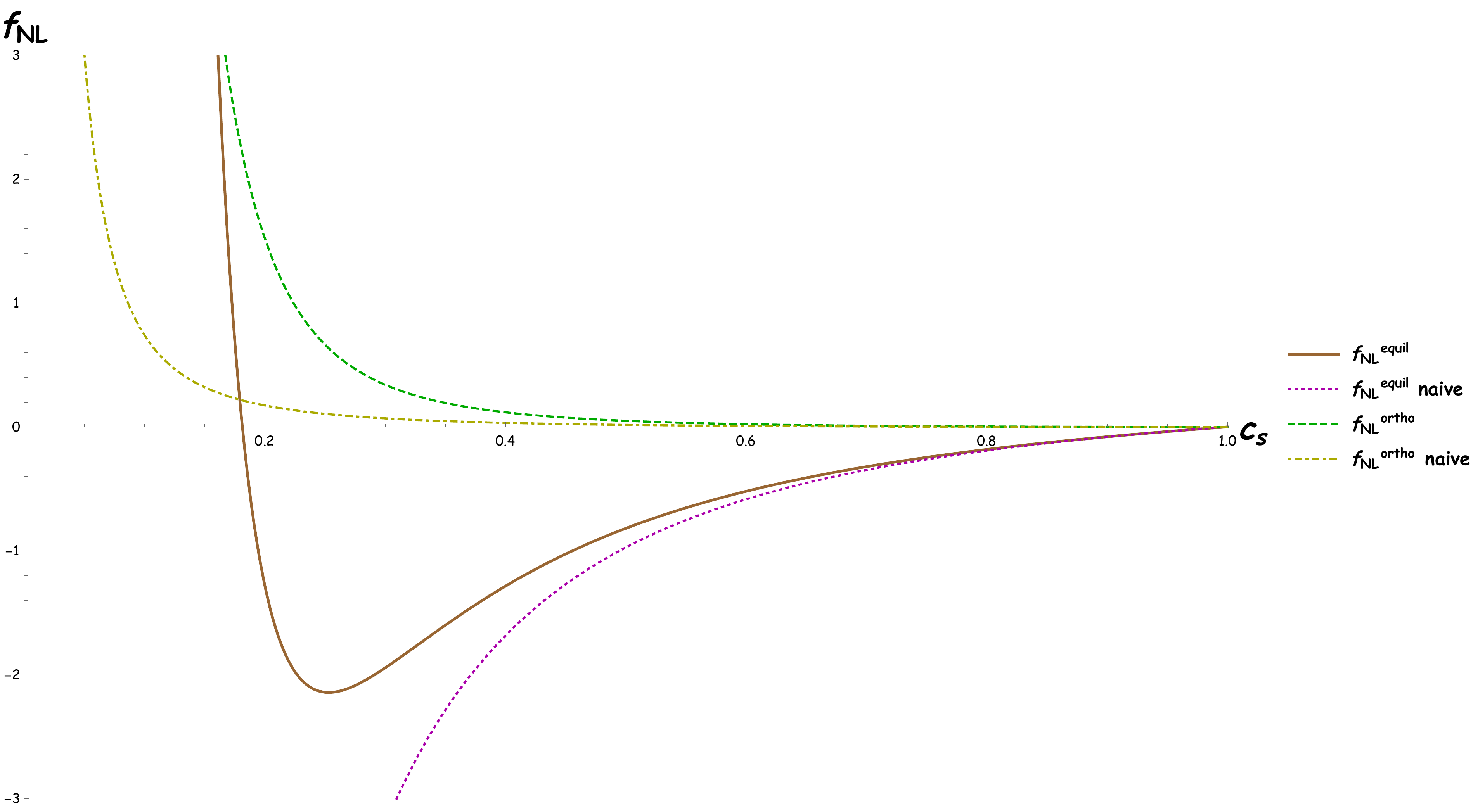}
      \caption{$f_\text{NL}^\text{equil}$, $f_\text{NL\,(naive)}^\text{equil}$, $f_\text{NL}^\text{ortho}$ and $f_\text{NL\,(naive)}^\text{ortho}$ vs. $c_s$.}
      \label{fig:fnlobsvscs}
\end{center}
\end{figure}

\noindent With $M_2$ and $M_3$ as given by \eqref{M2} and \eqref{M3} respectively, we find that \footnote{If the $V_{\sigma\sigma\sigma}$ term is neglected, $\mathcal{A} = -\frac{1}{2}(1 + c_s^2) \to f_\text{NL}^{\dot{\pi}^3} \sim -\frac{1}{2}\frac{(1 - c_s^4)}{c_s^2}$. Some authors \cite{Senatore:2009gt,Senatore:2010jy} argue that in order not to have an unnatural hierarchy between the scales associated with the two distinct operators $\dot{\pi}_c\,(\widetilde{\partial}_i\pi_c)^2$ and $\dot{\pi}_c^3$, one must require $\mathcal{A} \sim \mathcal{O}(1)$.}  
\begin{align}\label{Apar}
\mathcal{A} \equiv -c_s^2 + \frac{2}{3}\frac{M_3^4}{M_2^4}c_s^2 = -c_s^2 + c_s^2\left(\frac{1}{c_s^2} - 1\right)\left[\frac{R}{6M_\text{eff}^2}V_{\sigma\sigma\sigma} - \frac{1}{2}\right].
\end{align}
Using the minimum given by \eqref{min} and defining $\widetilde{M}_\text{eff}^2 \equiv M_\text{eff}^2\,c_s^{-2}$ we find through \eqref{speedandmass} and \eqref{Apar} that 
\begin{align}
c_s^2 &\simeq \frac{4\,\lambda\,v^2 - \dot{\theta}_0^2}{4\,\lambda\,v^2 + 3\,\dot{\theta}_0^2} \approxeq \frac{M_\text{eff}^2}{\widetilde{M}_\text{eff}^2},\label{cs2}\\
\mathcal{A} &\simeq \frac{ \dot{\theta}_0^4 + 8\,\lambda\,v^2\,\dot{\theta}_0^2 - 16\,\lambda^2\,v^4}{(4\,\lambda\,v^2 - \dot{\theta}_0^2)(4\,\lambda\,v^2 + 3\,\dot{\theta}_0^2)} \approxeq \frac{M_\text{eff}^2\,\widetilde{M}_\text{eff}^2 - 4(2\sqrt{2}\,\lambda\,v^2 - \dot{\theta}_0^2)(2\sqrt{2}\,\lambda\,v^2 + \dot{\theta}_0^2)}{M_\text{eff}^2\,\widetilde{M}_\text{eff}^2}.\label{estA} 
\end{align}
Then, using \eqref{cs2} and \eqref{estA} in \eqref{estfnltim} and \eqref{estfnlgrad}, we find the following estimates for the amplitude of non-Gaussianities
\begin{align}
&f_\text{NL}^{\dot{\pi}(\nabla\pi)^2} \sim \frac{4\,\dot{\theta}_0^2}{4\,\lambda\,v^2 - \dot{\theta}_0^2} \sim 4\frac{\dot{\theta}_0^2}{M_\text{eff}^2},\\
&f_\text{NL}^{\dot{\pi}^3} \sim \frac{4\,\dot{\theta}_0^2\,(\dot{\theta}_0^4 + 8\,\lambda\,v^2\,\dot{\theta}_0^2 - 16\,\lambda^2\,v^4)}{(4\,\lambda\,v^2 - \dot{\theta}_0^2)^2(4\,\lambda\,v^2 + 3\,\dot{\theta}_0^2)} \sim \frac{4\,\dot{\theta}_0^2}{M_\text{eff}^4\,\widetilde{M}_\text{eff}^2}\left[M_\text{eff}^2\,\widetilde{M}_\text{eff}^2 - 4(2\sqrt{2}\,\lambda\,v^2 - \dot{\theta}_0^2)(2\sqrt{2}\,\lambda\,v^2 + \dot{\theta}_0^2)\right].
\end{align}
The precise analysis using the so-called ``in-in'' formalism (see footnote \ref{inin}) gives the numerical coefficients we are missing for the exact prediction. With $f_\pi$ and $\Lambda$ defined through \eqref{fpi} and \eqref{Lambda} respectively, it can be shown that \cite{Baumann:2014nda} (see appendix \ref{nongaussdef})
\begin{align}
f_\text{NL}^{\dot{\pi}(\nabla\pi)^2} &= -\frac{85}{324}\left(\frac{f_\pi}{\Lambda}\right)^2 = -\frac{85}{81}\frac{\dot{\theta}_0^2}{M_\text{eff}^2} = -\frac{85}{324}\left(\frac{1}{c_s^2} - 1\right),\\
f_\text{NL}^{\dot{\pi}^3} &= +\frac{5\,\mathcal{A}}{81}\left(\frac{f_\pi}{\Lambda}\right)^2 = \frac{20}{81}\frac{\dot{\theta}_0^2}{M_\text{eff}^4\,\widetilde{M}_\text{eff}^2}\left[M_\text{eff}^2\,\widetilde{M}_\text{eff}^2 - 4(2\sqrt{2}\,\lambda\, v^2 - \dot{\theta}_0^2)(2\sqrt{2}\,\lambda\,v^2 + \dot{\theta}_0^2)\right]\nonumber\\
&= \frac{5}{81}\left[-\frac{5}{8} + \frac{1}{8\,c_s^4} - \frac{3}{8\,c_s^2} + \frac{7\,c_s^2}{8}\right],\label{fnlpidot3}
\end{align}
where, in order to get the last line in \eqref{fnlpidot3}, use has been made of the expression for $\dot{\theta}_0^2$ as given in \eqref{setofcond}, which also implies 
\begin{align}
M_\text{eff}^2 \approx \frac{16\,\lambda\,v^2\,c_s^2}{1 + 3\,c_s^2},\quad \widetilde{M}_\text{eff}^2 \approx \frac{16\,\lambda\,v^2}{1 + 3\,c_s^2} \quad \text{and} \quad \mathcal{A} = - \frac{1}{4} + \frac{1}{8\,c_s^2} - \frac{7\,c_s^2}{8}.\label{mmtanda}
\end{align}
We should compare the last expression in \eqref{fnlpidot3} with the naive prediction that one gets when using \eqref{coeffnaiveeft} in \eqref{Adeff} instead,
\begin{align}
f_\text{NL\,(naive)}^{\dot{\pi}^3} \equiv f_\text{NL}^{\dot{\pi}^3}\big|_{V_{\sigma\sigma\sigma} = 0} &= +\frac{5\,\mathcal{A}}{81}\left(\frac{f_\pi}{\Lambda}\right)^2\bigg|_{V_{\sigma\sigma\sigma} = 0} = \frac{5}{81}\left[-\frac{1}{2\,c_s^2} + \frac{c_s^2}{2}\right],\label{fnlnaive}
\end{align}
which is clearly negative when $c_s < 1$ and tends to $-\infty$ as $c_s$ decreases. The behavior of $f_\text{NL}^{\dot{\pi}^3}$, on the other hand, is quite different as can be anticipated by looking at \eqref{fnlpidot3}. Indeed, it possesses a zero-crossing point around $c_s \approx 0.51$, a global minimum around $c_s \approx 0.67$ (where $f_\text{NL}^{\dot{\pi}^3} \approx -2.76 \times 10^{-2}$) and tends to zero as $c_s$ approaches $1$, as it should. Also, due to the presence of the positive $c_s^{-4}$ term, $f_\text{NL}^{\dot{\pi}^3}$ tends to $+\infty$ as $c_s$ approaches zero.  All this can be seen in \figurename{\ref{fig:fnlpivscs}} where we plot $f_\text{NL}^{\dot{\pi}^3}$, $f_\text{NL\,(naive)}^{\dot{\pi}^3}$, $f_\text{NL}^{\dot{\pi}(\nabla\pi)^2}$ and $\mathcal{A}$ vs. $c_s$. Let us just comment that the scaling $f_\text{NL} \sim c_s^{-4}$ is not usual for non-canonical models like DBI \cite{Silverstein:2003hf,Alishahiha:2004eh} or $k$-inflation \cite{ArmendarizPicon:1999rj,Garriga:1999vw}, where it is a familiar result that $f_\text{NL} \sim c_s^{-2}$ \cite{Chen:2006nt}. This peculiar scaling does arise in Galileon models of inflation \cite{Burrage:2010cu} based on the so-called ``Galilean symmetry'' introduced in \cite{Nicolis:2008in} \footnote{In \cite{Burrage:2010cu} the $c_s^{-4}$ behavior appears since, due to symmetry, the dimension seven operator (after canonical normalization) $\partial^2\pi(\partial\pi)^2$ is naturally of comparable ``size'' with the usual $\dot{\pi}(\partial\pi)^2$ and, only for the latter, the non-linearly realized Lorentz invariance ``requires'' $f_\text{NL} \sim c_s^{-2}$.}.
\\
Planck 2015 \cite{Ade:2015ava} puts bounds on two specific linear combinations of $f_\text{NL}^{\dot{\pi}(\nabla\pi)^2}$ and $f_\text{NL}^{\dot{\pi}^3}$, namely the ``equilateral'' $f_\text{NL}^\text{equil}$ and the ``orthogonal'' $f_\text{NL}^\text{ortho}$ \footnote{$f_\text{NL}^\text{equil}$ and $f_\text{NL}^\text{ortho}$ are ``defined'' as the result of projecting the shapes associated with $f_\text{NL}^{\dot{\pi}(\nabla\pi)^2}$ and $f_\text{NL}^{\dot{\pi}^3}$ into the equilateral and orthogonal templates using the shape inner product introduced in \cite{Babich:2004gb}. For details see \cite{Senatore:2009gt}.}. The mean values of the estimators for $f_\text{NL}^\text{equil}$ and $f_\text{NL}^\text{ortho}$ are given by
\begin{align}
f_\text{NL}^\text{equil} &= \left[-\frac{11}{40} + \frac{39}{500}\,\mathcal{A}\right]\left(\frac{f_\pi}{\Lambda}\right)^2 = \frac{181}{800} + \frac{39}{4000\,c_s^4} - \frac{1217}{4000\,c_s^2} + \frac{273\,c_s^2}{4000},\label{fequi}\\
f_\text{NL}^\text{ortho} &= \left[\frac{159}{10000} + \frac{167}{10000}\,\mathcal{A}\right]\left(\frac{f_\pi}{\Lambda}\right)^2 = -\frac{2107}{80000} + \frac{167}{80000\,c_s^4} + \frac{771}{80000\,c_s^2} + \frac{1169\,c_s^2}{80000},\label{fort}
\end{align}
where we have used $\mathcal{A}$ as given in \eqref{mmtanda}. These are the ``physical'' constrained amplitudes of interest. In \figurename{\ref{fig:fnlobsvscs}} we plot $f_\text{NL}^\text{equil}$ and $f_\text{NL}^\text{ortho}$ vs. $c_s$ using the last expressions in \eqref{fequi} and \eqref{fort} along with the naive result of using $\mathcal{A}_\text{naive} = -\frac{1}{2}(1 + c_s^2)$ instead of $\mathcal{A}$.
\\
We observe again, that due to the presence of the $c_s^{-4}$ term, the behavior of the $f_\text{NL}$'s is quite different from the naive expectation when ignoring the non-linear self-interactions of the heavy field. In particular, we see that since
\begin{align}
f_\text{NL\,(naive)}^\text{equil} &\equiv f_\text{NL}^\text{equil}\big|_{V_{\sigma\sigma\sigma} = 0} = \left[-\frac{11}{40} + \frac{39}{500}\,\mathcal{A}\right]\left(\frac{f_\pi}{\Lambda}\right)^2\bigg|_{V_{\sigma\sigma\sigma} = 0} = \frac{11}{40} - \frac{157}{500\,c_s^2} + \frac{39\,c_s^2}{1000},\\
f_\text{NL\,(naive)}^\text{ortho} &\equiv f_\text{NL}^\text{ortho}\big|_{V_{\sigma\sigma\sigma} = 0} = \left[\frac{159}{10000} + \frac{167}{10000}\,\mathcal{A}\right]\left(\frac{f_\pi}{\Lambda}\right)^2\bigg|_{V_{\sigma\sigma\sigma} = 0} = -\frac{159}{10000} + \frac{151}{20000\,c_s^2} + \frac{167\,c_s^2}{20000},
\end{align}
$f_\text{NL\,(naive)}^\text{equil}$ ($f_\text{NL\,(naive)}^\text{ortho}$) tends to $-\infty$ ($+\infty$) as $c_s$ decreases. $f_\text{NL\,(naive)}^\text{equil}$ does not have a global minimum while $f_\text{NL}^\text{equil}$ does have one around $c_s \approx 0.25$ (where $f_\text{NL}^\text{equil} \approx -2.14$) and tends to $+\infty$ as $c_s$ decreases with a zero-crossing point around $c_s \approx 0.18$. $f_\text{NL}^\text{ortho}$ ($f_\text{NL\,(naive)}^\text{ortho}$) tends to $+\infty$ as $c_s^{-4}$ ($c_s^{-2}$) for small $c_s$ but otherwise stays very close to zero all the way up to $c_s = 1$, having a zero-crossing point around $c_s \approx 0.98$ ($c_s \approx 0.95$) and a global minimum at $c_s \approx 0.99$ ($c_s \approx 0.98$) where $f_\text{NL}^\text{ortho} \approx -9.79\times10^{-6}$ ($f_\text{NL\,(naive)}^\text{ortho} \approx -2.01\times10^{-5}$). We summarize the values of the different $f_\text{NL}$'s (for the same $c_s$'s that we considered in \figurename{\ref{fig:rvsns}} and Table \ref{Table1}) in Table \ref{Table2} below.
\begin{table}[h!]
\begin{center}
\begin{tabular}{|c|c|c|c|c|}
\cline{2-5}
\multicolumn{1}{c|}{} & $\bf{f}_\text{\,\bf{NL}}^\text{\bf{\,equil}}$ & $\bf{f}_\text{\bf{\,NL\,(naive)}}^\text{\bf{\,equil}}$ & $\bf{f}_\text{\bf{\,NL}}^\text{\bf{\,ortho}}$ & $\bf{f}_\text{\bf{\,NL\,(naive)}}^\text{\bf{\,ortho}}$ \\
\hline
$\bf{\color{red}{c_s = 0.999}}$ & $-7.067\times10^{-4}$ & $-7.069\times10^{-4}$ & $-1.536\times10^{-6}$ & $-1.569\times10^{-6}$ \\
\hline
$\bf{\color{ForestGreen}{c_s = 0.9}}$ & $-7.922\times10^{-2}$ & $-8.106\times10^{-2}$ & $5.785\times10^{-4}$ & $1.845\times10^{-4}$ \\
\hline
$\bf{\color{magenta}{c_s = 0.75}}$ & $-2.454\times10^{-1}$ & $-2.613\times10^{-1}$ & $5.613\times10^{-3}$ & $2.219\times10^{-3}$ \\
\hline
$\bf{\color{orange}{c_s = 0.47}}$ & $-9.362\times10^{-1}$ & $-1.138$ & $6.330\times10^{-2}$ & $2.012\times10^{-2}$ \\
\hline
\end{tabular}
\end{center}
\caption{$f_\text{NL}^\text{equil}$, $f_\text{NL\,(naive)}^\text{equil}$, $f_\text{NL}^\text{ortho}$ and $f_\text{NL\,(naive)}^\text{ortho}$ for different values of $c_s$.}
\label{Table2}
\end{table}
\\
The current Planck constraints at $2\sigma$ are \cite{Ade:2015ava}
\begin{align}
-156 < f_\text{NL}^\text{equil} &< 124, \quad& -100 < f_\text{NL}^\text{ortho} &< 32 \quad \text{(temperature data only)},\nonumber\\
-90 < f_\text{NL}^\text{equil} &< 82, \quad& -68 < f_\text{NL}^\text{ortho} &< 16 \quad \text{(temperature + polarization data)}.
\end{align}
Looking at Table \ref{Table2} we see that current observations are not sensitive enough to rule out the equilateral and orthogonal non-Gaussianities of our model. Needless to say, probing non-Gaussianities down to $f_\text{NL} \sim \mathcal{O}(1)$ or smaller is an important target for future experiments.\\
For completeness, let us mention that the so-called ``local'' shape with size $f_\text{NL}^\text{local}$ is much more well constrained (see \cite{Ade:2015ava} and appendix \ref{nongaussdef} for a definition of the local bispectrum). At $2\sigma$ Planck found that \cite{Ade:2015ava}
\begin{align}
-8.9 < f_\text{NL}^\text{local} &< 13.9, \text{(temperature data only)},\nonumber\\
-9.2 < f_\text{NL}^\text{local} &< 10.8, \text{(temperature + polarization data)}.
\end{align}
In \cite{Achucarro:2012sm} the EFT for single-field inflationary models descending from a ``parent theory'' containing several scalar fields was derived. Besides the cubic operators $\dot{\pi}^3$ and $\dot{\pi}(\nabla\pi)^2$ that we have found within our approximations, the following two terms were found in the ``decoupling limit'' (see appendix \ref{eftofinf} for a definition)
\begin{align}
S_\pi \ni \int d^4x\,a^3M_\text{Pl}^2\dot{H}\bigg(2\,\frac{\dot{c}_s}{c_s^3}\,\pi\,\dot{\pi}^2 + 2H\eta_\parallel\,\pi\left[\frac{\dot{\pi}^2}{c_s^2} - \frac{(\nabla\pi)^2}{a^2}\right]\bigg),
\end{align}
where $\eta_\parallel \equiv -\frac{\ddot{\varphi_0}}{H\dot{\varphi}_0}$ and $\varphi_0 \equiv R\,\theta_0$. These two operators lead to non-Gaussianities that satisfy the so-called Maldacena's consistency relation \cite{Maldacena:2002vr} in the sense that $f_\text{NL}^\text{local} \sim \mathcal{O}(\epsilon, \eta)$, confirming the fact that the $M_\text{eff}^2 \gg H^2$ limit is indeed a single-field scenario. In other words, even if the constraints in the local subcase are tighter, local non-Gaussianities are negligible in the $M_\text{eff}^2 \gg H^2$ limit, in agreement with the equivalence principle (see \cite{Creminelli:2004yq} for a general discussion of these points).\\
Now we will consider the $M_\text{eff}^2 \sim H^2$ case, which is quite different from the $M_\text{eff}^2 \gg H^2$ one as the heavy field cannot be integrated out anymore.

\subsection{$M_\text{eff}^2 \sim H^2$ regime.}
\subsubsection{The single-field EFT breaks down.}\label{eftbreaks}
Let us come back to \eqref{S2} and \eqref{S3}. In our specific model, taking $\sigma_0 = \widehat{\sigma} = 0$ so $R$ is determined by the naive VEV given by \eqref{min}, we have
\begin{align}
S^{(2)}[g_0,\phi_0,\delta\phi] &= \int d^4x\,a^3\bigg(-\frac{1}{2}\tensor{g}{^\mu^\nu}\partial_\mu\varphi\partial_\nu\varphi - 2M^2\cos\left(\frac{2\varphi_0}{R}\right)\varphi^2 + \frac{2}{R}\,\dot{\varphi}_0\,\dot{\varphi}\,\mathcal{F}\label{pertact2}\nonumber\\
&- 4M^2\sin\left(\frac{2\varphi_0}{R}\right)\varphi\,\mathcal{F} - \frac{1}{2}\tensor{g}{^\mu^\nu}\partial_\mu\mathcal{F}\partial_\nu\mathcal{F} - \frac{1}{2}M_\text{eff}^2\,\mathcal{F}^2\bigg),\\
S^{(3)}[g_0,\phi_0,\delta\phi] &= \int d^4x\,a^3\bigg(- \frac{1}{R}(\tensor{g}{^\mu^\nu}\partial_\mu\varphi\partial_\nu\varphi)\,\mathcal{F} + \frac{1}{R^2}\,\dot{\varphi}_0\,\dot{\varphi}\,\mathcal{F}^2 + \frac{4}{3R}M^2\sin\left(\frac{2\varphi_0}{R}\right)\varphi^3\nonumber\\ 
&- \frac{2M^2}{R}\sin\left(\frac{2\varphi_0}{R}\right)\varphi\,\mathcal{F}^2 - \frac{4M^2}{R}\cos\left(\frac{2\varphi_0}{R}\right)\varphi^2\mathcal{F} - \lambda\,R\,\mathcal{F}^3\bigg)\label{pertact3},
\end{align} 
where $\varphi_0 \equiv R\,\theta_0$, $\varphi \equiv R\,\delta\theta$ and we have used the definition $\mathcal{F} \equiv N^a\delta\phi_a = \delta\sigma$ which holds as long as $\dot{\sigma}_0 = 0$ \footnote{The change in ``notation'' $\delta\sigma \to \mathcal{F}$ makes contact with the literature and also aims for notational clarity.}. In appendix \ref{efftheory} we review for completeness the general conditions under which we can integrate out the high frequency degrees of freedom to get an effective single field theory \cite{Achucarro:2012yr}. It is clear though that when $M_\text{eff} \sim H$ integrating out the heavy mode is not justified as the cosmological experiment actually probes exactly this energy scale regime. We then need to consider the dynamics of the isocurvature perturbation $\mathcal{F}$ and its influence on the correlation functions of the adiabatic mode $\varphi$. Thus, we are interested in the action 
\begin{align}
S^{(0)}[g_0,\phi_0,\delta\phi] &\equiv \int d^4x\,a^3\left(-\frac{1}{2}\tensor{g}{^\mu^\nu}\partial_\mu\varphi\partial_\nu\varphi - \frac{1}{2}\tensor{g}{^\mu^\nu}\partial_\mu\mathcal{F}\partial_\nu\mathcal{F} - \frac{1}{2}M_\text{eff}^2\,\mathcal{F}^2\right),\label{freeact}\\
S^{\,\text{int}}[g_0,\phi_0,\delta\phi] &\equiv \int d^4x\,a^3\left(2\,\dot{\theta}_0\,\dot{\varphi}\,\mathcal{F} - \sqrt{2}\,\lambda\,v\,\mathcal{F}^3 - \frac{1}{4}\,\lambda\,\mathcal{F}^4\right),\label{intaction}
\end{align}
where we have neglected $\mathcal{O}(\beta)$ terms since we are dealing with the theory of fluctuations, we have taken $R \approx \sqrt{2}\,v$ and we have included the fourth-order term $\mathscr{L}^{\,\text{int}} \supset -\frac{1}{4!}V_{\sigma\sigma\sigma\sigma}(\sigma_0,\theta_0)(\delta\sigma)^4$ which is also not suppressed by any slow-roll conditions. Note that among the rest of the interaction terms in \eqref{pertact3} we have also neglected the ``irrelevant'' operators $(\partial\,\varphi)^2\,\mathcal{F}$ and $\dot{\varphi}_0\,\dot{\varphi}\,\mathcal{F}^2$ as they are suppressed by $\left(\frac{H}{v}\right)$ and $\left(\frac{H}{v}\right)^2$, respectively, while keeping the ``relevant'' operator $\mathcal{F}^3$. This is consistent with the analysis made in the original ``vanilla'' QSF model where it has been emphasized that the only operator that can (in principle) make $f_\text{NL} \gg 1$ is exactly the cubic term $\mathcal{F}^3$ (see Tables 1 and 2 of \cite{Chen:2009zp} and the discussion therein). Note also that the operator $\dot{\theta}_0\,\dot{\varphi}\,\mathcal{F}$ in \eqref{intaction} is second order in field fluctuations but still we treat it as an interaction (mixing) term. This is crucial for the perturbative Hamiltonian analysis that we now briefly review.\\
Starting from the full action $S[g_0,\phi_0,\delta\phi]$ we define the canonical momenta $\pi_{\delta\phi} \equiv \frac{\delta S}{\delta\delta\dot{\phi}}$ as usual. Then we construct the Hamiltonian as $\mathscr{H} = \sum_{\delta\phi}\pi_{\delta\phi}\delta\dot{\phi} - \mathscr{L}$ where the $\delta\dot{\phi}$ are expressed in terms of the $\pi_{\delta\phi}$ and the $\delta\phi$. We now divide $\mathscr{H}$ into a free-field $\mathscr{H}^{(0)}$ and an interacting part $\mathscr{H}^{\,\text{int}}$ and replace the $\pi_{\delta\phi}$ by $\pi_{\delta\phi}^I$, satisfying Hamilton's equations of the free-field Hamiltonian, meaning $\delta\dot{\phi}_I = \frac{\delta\mathscr{H}^{(0)}}{\delta\pi_{\delta\phi}}\big|_{\pi_{\delta\phi} = \pi_{\delta\phi}^I}$. We finally use this last definition to get rid of the $\pi_{\delta\phi}^I$ in terms of the $\delta\phi$ and $\delta\dot{\phi}$ (see \cite{Weinberg:2005vy,Chen:2010xka} for more details). In the case at hand, the free and interaction Hamiltonian densities $\mathscr{H}^{(0)}$ and $\mathscr{H}^{\text{int}}$ are then respectively given by
\begin{align}
\mathscr{H}^{(0)} &\equiv \frac{a^3}{2}\left\{\dot{\varphi_I}^2 + \frac{(\nabla\varphi_I)^2}{a^2} + \dot{\mathcal{F}_I^2} + \frac{(\nabla\mathcal{F}_I)^2}{a^2} + \widetilde{M}_\text{eff}^2\,\mathcal{F}_I^2\right\},\label{freeham}\\
\mathscr{H}^{\,\text{int}} &\equiv \mathscr{H}_2^I + \mathscr{H}_3^I = a^3\left\{-2\,\dot{\theta}_0\,\dot{\varphi_I}\,\mathcal{F}_I + \sqrt{2}\,\lambda\,v\,\mathcal{F}_I^3 + \frac{1}{4}\,\lambda\,\mathcal{F}_I^4\right\},\label{intham}
\end{align}
where the ``$_I$'' subscript highlights the fact that we now deal with interaction picture fields and 
\begin{align}
\widetilde{M}_\text{eff}^2 \equiv V_{\sigma\sigma} + 3\,\dot{\theta}_0^2 = M_\text{eff}^2\,c_s^{-2},\label{hammass}
\end{align}
where use has been made of \eqref{speedandmass}. 
It is interesting to note that $\widetilde{M}_\text{eff}^2$ is nothing but the low-energy effective theory cut-off \eqref{effcutoff}. In \figurename{\ref{fig:feynrules}} below, we draw the ``Feynman rules'' associated with the interaction Hamiltonian \eqref{intham}.
\\
\begin{figure}[h!]
\begin{center}
      \includegraphics[width=0.4\textwidth] {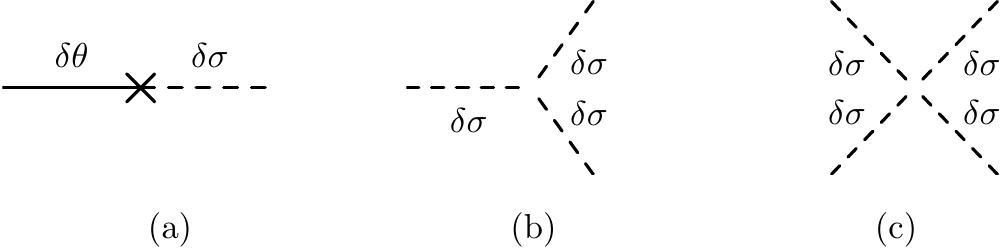}
      \caption{``Feynman rules'' for the interaction Hamiltonian $\mathscr{H}^{\,\text{int}}$. (a) is the so-called ``transfer function'' between adiabatic and isocurvature modes while (b) and (c) represent the three and four-point self-interaction terms of the isocurvature mode.}
      \label{fig:feynrules}
\end{center}
\end{figure}
\\
\noindent In order to rely on perturbation theory using $\mathscr{H}^{\text{int}}$ we will demand that \footnote{Let us emphasize that this perturbativity condition is \textit{not} tied to the Hamiltonian analysis. Within the so-called ``Schwinger-Keldysh'' formalism (SK), the generating functional $Z[J]$ is put into useful form by splitting the classical Lagrangian into free and interacting parts $\mathscr{L}[\phi] = \mathscr{L}_0[\phi] + \mathscr{L}_\text{int}[\phi]$, such that $Z[J] \sim \exp\left(i\int\mathscr{L}_\text{int}\left[\frac{\delta}{i\delta J}\right]\right)Z_0[J]$ where $Z_0[J] \sim \int\mathscr{D}\phi\exp\left(i\int\left\{\mathscr{L}_0[\phi] + J\phi\right\}\right)$. Since $Z_0[J]$ is a Gaussian integral, it can be carried out explicitly. Then the interaction piece is expanded \textit{perturbatively} to get the desired correlators. Barring unimportant subtleties, this is not different than good old QFT \`a la Feynman. See \cite{Chen:2017ryl} for a modern review of SK, its applicability on QSF inflation and original references.}
\begin{align}
\frac{\dot{\theta}_0^2}{H^2} &\ll 1\label{pertcond1}\\
&\text{and}\nonumber\\
\quad |V_{\sigma\sigma\sigma}|\,H^3 &\ll 3\,V_{\sigma\sigma}\,H^2.\label{pertcond2}
\end{align}
Condition \eqref{pertcond1} is necessary since the correction to the leading power spectrum is suppressed by the factor $\frac{\dot{\theta}_0^2}{H^2}$ (see \eqref{powersp} below). Condition \eqref{pertcond2} reflects the fact that, in the potential, the quadratic term should dominate over the cubic one when $\mathcal{F} \lesssim H$. In the QSF scenario, corresponding to $\widetilde{M}_\text{eff}^2 = V_{\sigma\sigma} + 3\,\dot{\theta}_0^2 \equiv \widetilde{\alpha}\,H^2$ with $\widetilde{\alpha} \sim \mathcal{O}(1)$, condition \eqref{pertcond2} is equivalent to
\begin{align}
\frac{|V_{\sigma\sigma\sigma}|}{H} \ll 3\,\widetilde{\alpha}\label{QSFsecpert}
\end{align}
as long as \eqref{pertcond1} simultaneously holds \cite{Chen:2009zp}. Within our model $V_{\sigma\sigma\sigma} = 6\,\lambda\,R \approx 6\sqrt{2}\,\lambda\,v$, so using a ``benchmark point'' compatible with \eqref{setofcond}, where we pick 
\begin{align}
v \approx 15.1\,M_\text{Pl} \approx 3.67 \times 10^{19}\,\text{GeV} \quad \text{and} \quad H \approx 6.6 \times 10^{13}\,\text{GeV},\label{benchmark}
\end{align}
condition \eqref{QSFsecpert} implies that $\lambda \ll (6.35 \times 10^{-7})\,\widetilde{\alpha}$. This last constraint on $\lambda$ is trivially satisfied since 
\begin{align}
\lambda \approx \frac{\widetilde{\alpha}\,H^2}{4\,v^2} \approx (8.08 \times 10^{-13})\,\widetilde{\alpha},
\end{align}
in agreement with the hierarchy $\dot{\theta}_0^2 \ll H^2 \ll v^2$.\\
Hamilton's equations deriving from the free Hamitonian \eqref{freeham} read
\begin{align}
\varphi_I'' + 2\,\mathcal{H}\,\varphi_I' + k^2\varphi_I &= 0,\label{eomvarphi}\\
\mathcal{F}_I'' + 2\,\mathcal{H}\,\mathcal{F}_I' + k^2\mathcal{F}_I + a^2\widetilde{M}_\text{eff}^2\,\mathcal{F}_I &= 0,\label{eomF}
\end{align}
where $f' \equiv \partial_\tau f$ and conformal time $\tau$ is defined through the relation $dt = a\,d\tau$, so in particular $\mathcal{H} \equiv \frac{a'}{a}$. Working in the de Sitter approximation ($\dot{H} = 0$) for simplicity \footnote{This is equivalent to neglect slow roll corrections to the ``Mukhanov-Sasaki'' equation \eqref{mukhsas}.} one finds that $\mathcal{H} = -\frac{1}{\tau}$ and $a = -\frac{1}{\mathcal{H}\,\tau}$. It is straightforward to show that if we define $u_k \equiv a\,\varphi_I$ and $v_k \equiv a\,\mathcal{F}_I$, the equations of motion \eqref{eomvarphi}-\eqref{eomF} can be put in the form
\begin{align}
y_k'' + \left(k^2 - \frac{\nu_y^2 - \frac{1}{4}}{\tau^2}\right)y_k = 0,\quad \nu_y^2 \equiv \frac{9}{4} - \frac{m_y^2}{H^2},\label{mukhsas}
\end{align}
where $m_y$ stands for the mass of the modes $y_k = \{u_k,v_k\}$. In the massless case, meaning $\nu_u = \frac{3}{2}$, the solutions to \eqref{mukhsas} are the so-called ``Bunch-Davies'' mode functions which are given by \cite{Bunch:1978yq}
\begin{align}
u_k(\tau) = \frac{H}{\sqrt{2k^3}}\left(1 + ik\tau\right)e^{-ik\tau}.\label{bunchdavies}
\end{align}
For $m_v \equiv \widetilde{M}_\text{eff} \neq 0$ the more general solutions to \eqref{mukhsas} need to be considered. These are given by \footnote{It is worth considering the behavior of the mode functions after horizon exit, namely, as $k\tau \to 0$.\\
When $\frac{\widetilde{M}_\text{eff}^2}{H^2} \leq \frac{9}{4}$
\begin{align}
v_k(\tau) \to 
\begin{cases}
-e^{i\left(\nu + \frac{1}{2}\right)\frac{\pi}{2}}\frac{2^{\nu - 1}}{\sqrt{\pi}}\Gamma(\nu)\frac{H}{k^\nu}(-\tau)^{-\nu + \frac{3}{2}},\quad\text{for}\quad 0 < \nu \leq \frac{3}{2},\label{solapprox1}\\
e^{i\frac{\pi}{4}}\frac{1}{\sqrt{\pi}}H(-\tau)^{\frac{3}{2}}\ln(-k\tau),\quad\text{for}\quad \nu = 0.
\end{cases}
\end{align}
When $\frac{\widetilde{M}_\text{eff}^2}{H^2} > \frac{9}{4}$
\begin{align}
v_k \to -ie^{-\frac{\pi}{2}\mu + i\frac{\pi}{4}}\frac{\sqrt{\pi}}{2}H(-\tau)^{\frac{3}{2}}\left[\frac{1}{\Gamma(i\mu + 1)}\left(\frac{-k\tau}{2}\right)^{i\mu} - i\frac{\Gamma(i\mu)}{\pi}\left(\frac{-k\tau}{2}\right)^{-i\mu}\right],\label{solapprox2}
\end{align}
where $\mu \equiv \sqrt{\frac{\widetilde{M}_\text{eff}^2}{H^2} - \frac{9}{4}}$.}
\begin{align}
v_k(\tau) = 
\begin{cases}\label{solsms}
-i\,e^{i\left(\nu + \frac{1}{2}\right)\frac{\pi}{2}}\frac{\sqrt{\pi}}{2}H(-\tau)^{\frac{3}{2}}\mathbb{H}_\nu^{(1)}(-k\tau),\quad\text{for}\quad\frac{\widetilde{M}_\text{eff}^2}{H^2} \leq \frac{9}{4},\\
-i\,e^{-\frac{\pi}{2}\mu + i\frac{\pi}{4}}\frac{\sqrt{\pi}}{2}H(-\tau)^{\frac{3}{2}}\mathbb{H}_{i\mu}^{(1)}(-k\tau),\quad\text{for}\quad\frac{\widetilde{M}_\text{eff}^2}{H^2} > \frac{9}{4},
\end{cases}
\end{align}
where $\mathbb{H}_\nu^{(1)}$ is the Hankel function of the first kind and $\mu \equiv \sqrt{\frac{\widetilde{M}_\text{eff}^2}{H^2} - \frac{9}{4}}$. The normalization of the mode functions are chosen so that when the physical momentum $\frac{k}{a}$ is much larger than the Hubble parameter $H$ and the mass $m_y$, we get back the Bunch-Davies vacuum, i.e., $u_k$ as given in \eqref{bunchdavies} and $v_k \approx i\frac{H}{\sqrt{2k}}\,\tau e^{-ik\tau}$. We see from \eqref{solapprox1} that when $0 \leq \widetilde{M}_\text{eff} \leq \frac{3}{2}H$ the amplitude of the mode $v_k$ decays as $(-\tau)^{\frac{3}{2} - \nu}$ after horizon exit, so the lighter the isocurvaton is, the slower it decays. In the limit $\widetilde{M}_\text{eff} \to 0$ $\left(\nu \to \frac{3}{2}\right)$ the amplitude is frozen. On the other hand, when $\widetilde{M}_\text{eff} > \frac{3}{2}H$, we see from \eqref{solapprox2} that $v_k$ not only contains a decay factor $(-\tau)^{\frac{3}{2}}$ but an oscillation factor $\tau^{\pm i\mu}$ as well. While this oscillation is marginal for $\widetilde{M}_\text{eff} \sim H$, it causes cancellations in the integrals of the correlation functions and is equivalent to factors of Boltzmann-like suppression $\sim e^{-\frac{\widetilde{M}_\text{eff}}{H}}$ in the $\widetilde{M}_\text{eff} \gg H$ limit \footnote{In analogy to thermal field theory, the contributions of massive states to correlation functions are exponentially suppressed by a Boltzmann factor if the mass is much higher than the temperature. In de Sitter space there is a ``Gibbons-Hawking'' temperature given by $T_\text{GH} = \frac{H}{2\pi}$ \cite{Gibbons:1977mu} and hence the corresponding Boltzmann factor reads $e^{-\frac{\widetilde{M}_\text{eff}}{T_\text{GH}}} = e^{-\frac{2\pi\widetilde{M}_\text{eff}}{H}}$.}. This is the reason behind the fact that most authors originally considered the $0 \leq \nu \leq \frac{3}{2}$ regime only.

\newpage

\begin{figure}[h]
\begin{center}
      \includegraphics[width=0.4\textwidth] {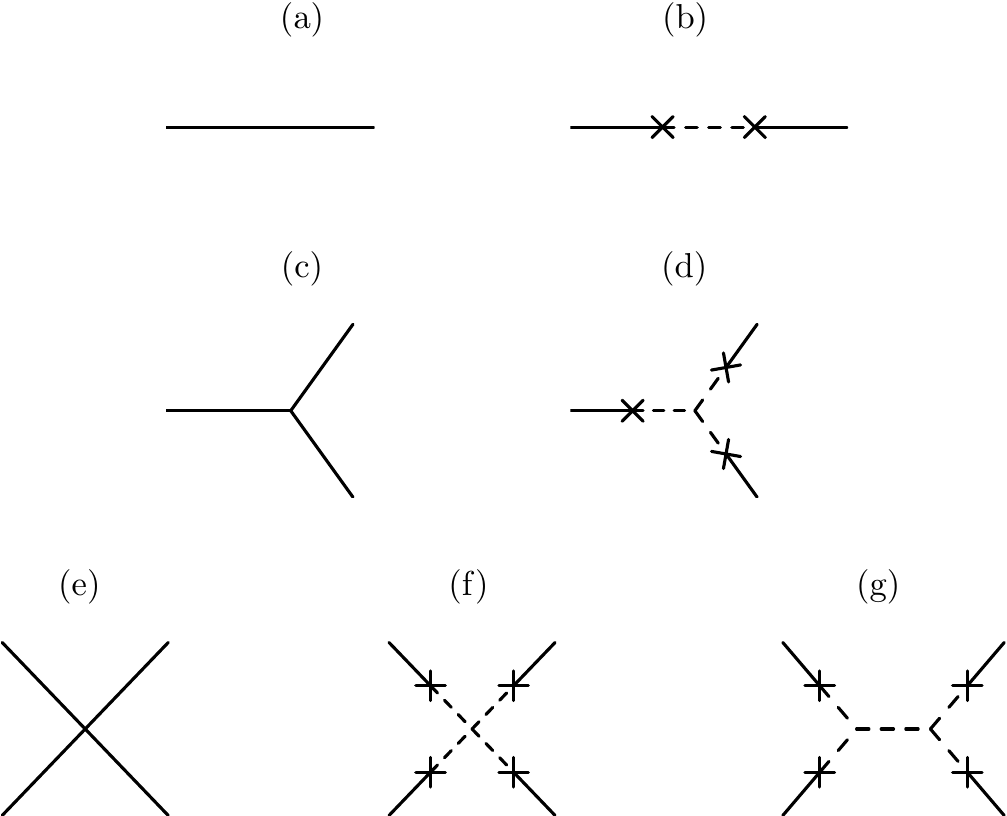}
      \caption{Leading (tree-level) diagrams for the 2, 3 and 4-point functions of the curvature perturbation in QSF inflation models. (a), (c) and (e) represent the naive (tree-level) correlators while (b), (d), (f) and (g) are the leading corrections through isocurvature (tree-level) mediation.}
      \label{fig:correlators}
\end{center}
\end{figure}

\noindent However it has been recently understood that the regime $\widetilde{M}_\text{eff} \gtrsim \frac{3}{2}H$ has very peculiar features in the so-called ``squeezed limit'' that however, realistically, will only be disentangled after finding some first evidence of non-Gaussianities \cite{Arkani-Hamed:2015bza,Meerburg:2016zdz} \footnote{The regime $\widetilde{M}_\text{eff} \gg H$ is not trivial. The time-dependent inflationary background implies that integrating out a heavy mode leaves an imprint in the speed of sound of adiabatic fluctuations, as we have discussed thoroughly in subsection \ref{eftadiab} and (in some generality) appendix \ref{efftheory}. See \cite{Achucarro:2012yr} and references therein.}. We are interested in the perturbative corrections to the 2, 3 and 4-point functions of the adiabatic fluctuation. In \figurename{\ref{fig:correlators}} we draw the (tree-level) correlators along with the perturbative corrections due to the presence of the isocurvature mode.
\\
The standard tool to calculate cosmological correlators is the in-in formalism \footnote{\label{inin}See \cite{Weinberg:2005vy} and references therein. For a highly improved covariant prescription to calculate cosmological correlators see the SK formalism recently presented in \cite{Chen:2017ryl}.}. The master formula of in-in applied to the two-point function of $\varphi$ is given by
\begin{align}
\langle\varphi^2\rangle = \bigg\langle0\bigg|\left[\overline{T}\,\text{exp}\left(i\int_{-\infty^-}^t dt''\,H_I(t'')\right)\right]\varphi^2_I(t)\left[T\,\text{exp}\left(-i\int_{-\infty^+}^t dt'\,H_I(t')\right)\right]\bigg|0\bigg\rangle,
\end{align}
where $H_I = \int d^3\bold{x}\,\mathscr{H}_2^I$, $\overline{T}$ is the anti-time-ordering symbol and $\infty^\pm \equiv \infty(1\pm i\,\varepsilon)$. Then, recalling that $\mathcal{R} \approx -\frac{H}{\dot{\varphi}_0}\varphi$, the dimensionless power spectrum of curvature fluctuations $\Delta_\mathcal{R}^2$ as defined in \eqref{dimpowspec} and the scalar tilt, defined as $n_s \equiv 1 + \frac{d\,\text{ln}\,\Delta_\mathcal{R}^2}{d\,\text{ln}\,k}$, are given by \footnote{In \cite{Pi:2012gf,Chen:2012ge} it was proven that, in the large effective mass limit, $\mathfrak{C}(\mu) \approx \frac{2}{\mu^2}$ with $\mu^2 \equiv \frac{\widetilde{M}_\text{eff}^2}{H^2} - \frac{9}{4} \approx \frac{\widetilde{M}_\text{eff}^2}{H^2}$. Using this in \eqref{powersp} we get that $\Delta_\mathcal{R}^2 \approx \mathring{\Delta}_\mathcal{R}^{2}\left(1 + 2\frac{\dot{\theta}_0^2}{\widetilde{M}_\text{eff}^2}\right)$, where $\mathring{\Delta}_\mathcal{R}^{2}$ stands for the single field ($c_s = 1$) power spectrum. This should be compared with the effective single field ($c_s \neq 1$) prediction which in this case is given by $\Delta_\mathcal{R}^{2} \approx \mathring{\Delta}_\mathcal{R}^{2}\,c_s^{-1} \approx \mathring{\Delta}_\mathcal{R}^{2}\left(1 + 2\frac{\dot{\theta}_0^2}{M_\text{eff}^2}\right) = \mathring{\Delta}_\mathcal{R}^{2}\left(1 + 2\frac{\dot{\theta}_0^2}{\widetilde{M}_\text{eff}^2}\,c_s^{-2}\right)$, so both predictions coincide to $\mathcal{O}\left(\frac{\dot{\theta}_0^2}{\widetilde{M}_\text{eff}^2}\right)$.}
\begin{align}
\Delta_\mathcal{R}^2 = \frac{H^4}{4\pi^2\dot{\varphi}_0^2}\left[1 + \mathfrak{C}(\nu)\left(\frac{\dot{\theta}_0}{H}\right)^2\right], \quad n_s - 1 = -2\,\epsilon - \eta + \eta\,\mathfrak{C}(\nu)\left(\frac{\dot{\theta}_0}{H}\right)^2.\label{powersp}
\end{align}
The explicit analytic calculation of $\mathfrak{C}(\nu)$ for arbitrary $\widetilde{M}_\text{eff}$ can be found in \cite{Chen:2012ge} (see \cite{Chen:2017ryl} for a ``quick'' derivation). The result in \eqref{powersp} justifies the necessity of the first perturbative condition in \eqref{pertcond1} \footnote{Strictly speaking, the perturbative condition one needs to impose is $\mathfrak{C}(\nu)\,\dot{\theta}_0^2 \ll H^2$. $\mathfrak{C}(\nu)$ is a very slowly growing function that stays $\mathcal{O}(1)$ until it diverges as $\nu\to\frac{3}{2}$, as can be seen in FIG.6 of reference \cite{Chen:2009zp} and equation (3.15) of reference \cite{Assassi:2012zq}. The divergence represents the massless limit, and since we want to focus in the regime where $\widetilde{M}_\text{eff}^2 \gtrsim H^2$, we can safely take $\mathfrak{C}(\nu) \sim \mathcal{O}(1)$ for all our purposes.}.

\subsubsection{Non-Gaussianities.}
As for the bispectrum it can be shown \cite{Chen:2009zp} that in the squeezed limit $p_3 \ll p_1 \simeq p_2$, when $\nu \neq 0$ $\left(0 \leq \widetilde{M}_\text{eff} < \frac{3}{2}H\right)$, the curvature scalar bispectrum that one gets through the $V_{\sigma\sigma\sigma}$ interaction of the isocurvature mode has a momentum dependence that scales as
\begin{align}
\langle\mathcal{R}_{\bold{p}_1}\mathcal{R}_{\bold{p}_2}\mathcal{R}_{\bold{p}_3}\rangle \sim \frac{1}{p_1^3\,p_3^3}\left(\frac{p_3}{p_1}\right)^{\frac{3}{2} - \nu} \quad \text{when} \quad p_3 \ll p_1 \simeq p_2 \quad \text{and} \quad \nu \neq 0.\label{momdep}
\end{align}
Looking at \eqref{solsms} we can understand this momentum dependence by recalling that under Bunch-Davies initial conditions a correlation between long and short wavelengths can only be generated once the short wavelength modes approach horizon scales. The amplitude of the long wavelength will have decayed according to the factor $\left(\frac{\tau_1}{\tau_3}\right)^{\frac{3}{2} -\nu} = \left(\frac{p_3}{p_1}\right)^{\frac{3}{2} -\nu}$ by that time, explaining the behavior in \eqref{momdep}. The shape function \eqref{momdep} has been dubbed ``intermediate'' \cite{Chen:2009we} since it interpolates between local and equilateral shapes as $\nu \to \left\{\frac{3}{2},\,0\right\}$, respectively (see \cite{Chen:2010xka} for standard definitions). Indeed, the more massive the isocurvature mode is, the faster it decays on super-horizon scales, so the largest contribution to non-Gaussianities is generated around horizon-crossing scales, i.e. in the equilateral configuration.

\newpage

\noindent On the other hand, if the isocurvature mode is lighter, the super-horizon isocurvature fluctuations survive longer and can contribute to correlations between long and short modes, i.e. in the so-called local configuration.\\
We can estimate the size of non-Gaussianities, i.e. the order of magnitude of $f_\text{NL}$, by realizing that the dimensionless coupling constants for the cubic isocurvature interaction and the transfer vertex go like $\left(\frac{V_{\sigma\sigma\sigma}}{H}\right)$ and $\left(\frac{\dot{\theta}_0}{H}\right)$, respectively \cite{Chen:2009zp}. Thus, since $\mathcal{R} \sim \sqrt{\Delta_\mathcal{R}^2}$, we find through inspection of diagram (d) in \figurename{\ref{fig:correlators}} that
\begin{align}
\left\langle\mathcal{R}^3\right\rangle \sim \left(\frac{V_{\sigma\sigma\sigma}}{H}\right)\left(\frac{\dot{\theta}_0}{H}\right)^3(\Delta_\mathcal{R}^{2})^{3/2} \sim f_\text{NL}\,(\Delta_\mathcal{R}^{2})^2 \to f_\text{NL} \sim \frac{1}{\sqrt{\Delta_\mathcal{R}^{2}}}\left(\frac{V_{\sigma\sigma\sigma}}{H}\right)\left(\frac{\dot{\theta}_0}{H}\right)^3.\label{fnlest}
\end{align}
\\
In our model, $V_{\sigma\sigma\sigma} = 6\,\lambda\,R \approx 6\sqrt{2}\,\lambda\,v$ and $\lambda \approx \frac{\widetilde{\alpha}\,H^2}{4\,v^2}$ thus $\frac{V_{\sigma\sigma\sigma}}{H} \approx \frac{3\sqrt{2}\,\widetilde{\alpha}}{2}\frac{H}{v} \approx (3.81 \times 10^{-6})\,\widetilde{\alpha}$, where we have used the benchmark point defined in \eqref{benchmark}. Taking $\sqrt{\Delta_\mathcal{R}^{2}} \approx 4.63 \times 10^{-5}$ (from observations) 
\begin{align}
f_\text{NL} \sim (8.23 \times 10^{-2})\,\widetilde{\alpha}\left(\frac{\dot{\theta}_0}{H}\right)^3.\label{ormagfnl1}
\end{align}
If we assume a non-conservative value $\frac{\dot{\theta}_0}{H} \approx \frac{1}{\sqrt{10}}$ (so we get an $\mathcal{O}(1  0^{-1})$ correction to the power spectrum in \eqref{powersp}) we find using \eqref{ormagfnl1} that
\begin{align}
f_\text{NL} \sim (2.6 \times 10^{-3})\,\widetilde{\alpha}.\label{ormagfnl2}
\end{align}
The estimation above lacks a numerical factor (and a sign) that Chen and Wang originally obtained. Quoting their result,
\begin{align}
f_\text{NL} \approx \frac{\vartheta(\nu)}{\sqrt{\Delta_\mathcal{R}^{2}}}\left(\frac{-V_{\sigma\sigma\sigma}}{H}\right)\left(\frac{\dot{\theta}_0}{H}\right)^3,\label{fnlbettest}
\end{align}
where $\vartheta(\nu)$ is a positive numerical coefficient which is expected to be $\mathcal{O}(1)$ \footnote{\label{infracut}It can be numerically shown that $\vartheta(\nu)$ blows up as $\nu \to \frac{3}{2}$ ($\widetilde{M}_\text{eff} \to 0$). The divergence occurs because we use the constant turn assumption. However when $\widetilde{M}_\text{eff} = 0$, a $\delta\sigma$ fluctuation never decays at super-horizon so the transfer from isocurvaton to curvaton lasts forever. As \cite{Chen:2009zp} points out, if the horizon crossing time of a perturbation mode is $N_f$ e-folds before the end of inflation (or the time when the inflaton trajectory becomes straight), one needs to impose a cut-off in the conformal time integrals of the exact in-in formula for $\langle\mathcal{R}^3\rangle$. All in all we could naively conclude that the integrals are dominated by a $N_f^4$ behavior. However, we need to realize that in this limit, $\mathfrak{C}(\nu)$ in \eqref{powersp} scales as $N_f^2$ for the same reason. For large $N_f$, the perturbativity condition becomes $N_f^2\left(\frac{\dot{\theta}_0^2}{H^2}\right) \ll 1$ instead. Thus, in the perturbative regime, the effective ``enchancement'' factor is only $N_f$ (which in principle can be as large as 60). Since we are not interested in the ``multifield'' inflation limit \cite{Senatore:2010wk}, $\vartheta(\nu)$ is $\mathcal{O}(1)$ for our purposes.}.

\newpage

\noindent Then our estimation \eqref{ormagfnl1} is slightly modified to finally give 
\begin{align}
f_\text{NL} \approx -(2.6 \times 10^{-3})\,\widetilde{\alpha}\,\vartheta(\nu)\,\lesssim\,\mathcal{O}(\epsilon,\eta),\quad \left(\widetilde{\alpha},\,\vartheta(\nu) \sim \mathcal{O}(1)\,\text{numbers}\right)
\end{align}
which is (still) unobservably small.\\
Finally, we can estimate the trispectra $\tau_\text{NL}$ (4-point function) by considering diagrams (f) and (g) in \figurename{\ref{fig:correlators}}. We get that
\begin{align}
\tau_\text{NL} \sim \text{max}\left\{\tau_\text{NL}^\text{SE} \cong \frac{1}{\Delta_\mathcal{R}^{2}}\left(\frac{\dot{\theta}_0}{H}\right)^4\left(\frac{V_{\sigma\sigma\sigma}}{H}\right)^2,\,\tau_\text{NL}^\text{CI} \cong \frac{1}{\Delta_\mathcal{R}^{2}}\left(\frac{\dot{\theta}_0}{H}\right)^4V_{\sigma\sigma\sigma\sigma}\right\}\label{tnlest},
\end{align}
where, following \cite{Chen:2009zp}, SE and CI in $\tau_\text{NL}^\text{SE}$ and $\tau_\text{NL}^\text{CI}$ stand for ``scalar-exchange'' and ``contact-interaction'', respectively.
Recalling that in our model, $V_{\sigma\sigma\sigma\sigma} = 6\,\lambda \approx \frac{3}{2}\,\widetilde{\alpha}\left(\frac{H}{v}\right)^2$, we find that
\begin{align}
\tau_\text{NL} \sim \text{max}\,\left\{\tau_\text{NL}^\text{SE} \sim 6.78 \times 10^{-3}\,\widetilde{\alpha}^2\left(\frac{\dot{\theta}_0}{H}\right)^4,\, \tau_\text{NL}^\text{CI} \sim 2.27 \times 10^{-3}\,\widetilde{\alpha}\left(\frac{\dot{\theta}_0}{H}\right)^4\right\}.\label{taumax1}
\end{align}
Assuming again that $\frac{\dot{\theta}_0}{H} \approx \frac{1}{\sqrt{10}}$ this becomes
\begin{align}
\tau_\text{NL} \sim \text{max}\,\left\{\tau_\text{NL}^\text{SE} \sim 6.78 \times 10^{-5}\,\widetilde{\alpha}^2,\, \tau_\text{NL}^\text{CI} \sim 2.27 \times 10^{-5}\,\widetilde{\alpha}\right\}.\label{taumax2}
\end{align}
Considering \eqref{fnlest} and \eqref{tnlest} we see that 
\begin{align}
\tau_\text{NL}^\text{SE} \sim \left(\frac{H}{\dot{\theta}_0}\right)^2 f_\text{NL}^2 \quad \text{and} \quad \tau_\text{NL}^\text{CI} \sim \left(\frac{H}{\dot{\theta}_0}\right)^2\left(\frac{V_{\sigma\sigma\sigma\sigma}\,H^2} {{V_{\sigma\sigma\sigma}}^2}\right) f_\text{NL}^2.\label{taufnlsq}
\end{align}
As as consequence of perturbativity, we find that 
\begin{align}
\tau_\text{NL}^\text{SE} \gg \left(\frac{6}{5}f_\text{NL}\right)^2,
\end{align}
so that the so-called ``Suyama-Yamaguchi bound'' \cite{Suyama:2007bg} is satisfied as expected in the QSF scenario \cite{Assassi:2012zq} \footnote{The Suyama-Yamaguchi bound reads
\begin{align}
\tau_\text{NL}^\text{SE} \geq \left(\frac{6}{5}f_\text{NL}\right)^2.\label{suyyam}
\end{align}
The inequality is saturated for single-field inflation while multifield inflation satisfies \eqref{suyyam}. The case for QSF is in principle distinguishable as $\tau_\text{NL}^\text{SE} \gg \left(\frac{6}{5}f_\text{NL}\right)^2$ is expected to hold instead. See \cite{Assassi:2012zq} for a discussion.}. 

\newpage

\noindent We also see that for our specific model \eqref{taufnlsq} implies that
\begin{align}
\tau_\text{NL}^\text{CI} \sim \left(\frac{H}{\dot{\theta}_0}\right)^2\left(\frac{1}{3\,\widetilde{\alpha}}\right)f_\text{NL}^2 \quad \to \quad \tau_\text{NL}^\text{SE} > \tau_\text{NL}^\text{CI} \quad \text{when} \quad \widetilde{\alpha} \sim \mathcal{O}(1),
\end{align}
which is the case in \eqref{taumax1}. Interestingly, this ``hierarchy'' reverses when $\widetilde{\alpha} \leq \frac{1}{3}$. This fact could, in principle, be used to pin down the mass range of the isocurvatons of the QSF scenario once the ``Cosmological Collider Physics'' program is up and running \cite{Arkani-Hamed:2015bza}\cite{Meerburg:2016zdz}. Needless to say, measuring the trispectra of primordial density perturbations is way beyond our current experimental expectations.

\subsubsection{Comments on the $M_\text{eff}^2 \ll H^2$ regime.}\label{commlightmass}
As has been previously stressed, when approaching the isocurvaton light mass limit, the squeezed limit of the bispectrum in the QSF scenario is of ``quasi-local'' type, and the fluctuations decay much slower than in the heavy mass case. This situation has been originally discussed in \cite{Chen:2009zp}, where the following two instances have been distinguished:
\begin{itemize}
\item If $V_{\sigma\sigma\sigma}$ is still ``large'', the QSF analysis does apply, so we can use \eqref{fnlbettest} to estimate the size of non-Gaussianities, but with an infrared e-folds cutoff as discussed in footnote \ref{infracut}.
\item It is possible that in this limit the isocurvature background solution slow-rolls as well as the inflationary one, implying through slow-roll conditions that the coupling $\left(\frac{V_{\sigma\sigma\sigma}}{H}\right)_{\text{sr}}$ is $\sim \mathcal{O}\left(\epsilon^{3/2}\right)\frac{H}{M_\text{Pl}}$. As is well known \cite{Vernizzi:2006ve}, this scenario does not produce sizable non-Gaussianities.
\end{itemize}
Let us then analyze the isocurvaton light mass limit of our model to see into which of the above cases it falls. The light mass condition, $\widetilde{M}_\text{eff}^2 \approx 4\,\lambda\,v^2 + 3\,\dot{\theta}_0^2 \ll H^2$, amounts to $\lambda \ll \frac{H^2}{4\,v^2}$, as $\frac{\dot{\theta}_0^2}{H^2} \ll 1$ due to perturbativity.  Using the benchmark point \eqref{benchmark}, which is required by the background NI theory, this implies that $\lambda \ll 8.08 \times10^{-13}$. Since $V_{\sigma\sigma\sigma} = 6\sqrt{2}\,\lambda\,v$, we find that $\frac{V_{\sigma\sigma\sigma}}{H} \ll 3.81 \times 10^{-6}$. On the other hand, assuming $\epsilon \sim 10^{-2}$, we see that this last constraint on $\frac{V_{\sigma\sigma\sigma}}{H}$ takes us quite close to the slow-roll regime as $\left(\frac{V_{\sigma\sigma\sigma}}{H}\right)_{\text{sr}} \sim \epsilon^{3/2}\frac{H}{M_\text{Pl}} \sim 2.72 \times 10^{-8}$. We then realize that due to the tight symmetry constraints on the parameters of our model, non-Gaussianities are much more suppressed in the isocurvaton light mass scenario when compared to the QSF regime ones, which are already quite small. For this reason we do not further discuss this particular limit.

\section{Discussion and Conclusions.}\label{discandconc}
We have considered a generalization of Natural Inflation \cite{Freese:1990rb,Adams:1992bn} where the dynamics of the radial mode $\sigma$ is included. To this end we have carried out an educated field-theoretic construction of a ``UV-complete'' two-field model undergoing spontaneous as well as explicit symmetry breaking of a global $U(1)$ symmetry. The (soft) explicit symmetry breaking operators of our model give the (inflaton) pseudo-Goldstone field $\theta$ a naturally small mass in accordance with slow-roll requirements and makes the potential for the two-field system $V(\sigma,\theta)$ non-separable. We analyzed the dynamics of the background solution assuming an almost constant angular speed circular motion in (flat) field space. As for the theory of fluctuations, the results depend crucially on whether the effective mass squared of the radial field $M_\text{eff}^2$ is very heavy ($\gg$) or not ($\sim$) with respect to the cosmological collider experiment energy scale squared, $H^2$.
\\
\indent We have found that effective single-field Natural Inflation ($M_\text{eff}^2 \gg H^2$) has a better fit to current bounds in the $(n_s, r)$ plane \cite{Ade:2015lrj} if the speed of sound of adiabatic fluctuations $c_s$ is mildly smaller than one \footnote{See \cite{Achucarro:2015rfa} for previous developments along these lines.}. However the amplitudes of non-Gaussianities, collectively denoted as $f_\text{NL}$, are negligible unless $c_s^2 \ll 1$. In particular, we have noticed that the assumptions on the relative ``weight'' of the heavy field operators when neglecting its dynamics changes the behavior of $f_\text{NL}$ as a function of $c_s$, especially in the small $c_s$ regime. Indeed, keeping the $V_{\sigma\sigma\sigma}$ contribution in the constraint equation for $\delta\sigma$ changes the predictions of the model quite dramatically, as was argued in \cite{Gong:2013sma}. This ``free parameter'' (from the single-field EFT of inflation \cite{Cheung:2007st} point of view) is constrained by the symmetry or our model and feeds into the functional dependence of $f_\text{NL} = f_\text{NL}(c_s)$ leaving a characteristic behavior \footnote{Indeed our model is quite peculiar in the sense that in \eqref{Apar}, all terms are related by symmetry in such a way that $\frac{R}{6M_\text{eff}^2}V_{\sigma\sigma\sigma} \approx \frac{3}{8} + \frac{1}{8}c_s^{-2}$, so $f_\text{NL} = f_\text{NL}(c_s)$ ultimately. This kind of simplification does not occur in a generic model.}, that in the small $c_s$ regime, scales like $f_\text{NL} \sim c_s^{-4}$ instead of the usual $f_\text{NL} \sim c_s^{-2}$ scaling that is naively expected in this class of models \cite{Chen:2006nt,Ade:2015ava} (the $f_\text{NL} \sim c_s^{-4}$ scaling does arise, for example, in Galileon models of infation \cite{Burrage:2010cu} \footnote{It would be interesting to clarify the connection between such a scenario and the case where we integrate out the radial mode of a pseudo-Goldstone model without neglecting its self-interactions. This, however, lied beyond the scope of this paper.}). In our model, to get small $c_s$ such that the $f_\text{NL}$'s get any chance of being observable, requires a bit of tuning of initial conditions which is obviously unappealing from the theoretical point of view.
\\
\indent The other possibility that we have analyzed is the $M_\text{eff}^2 \sim H^2$ scenario, i.e. the Quasi-Single-Field regime \cite{Chen:2009we,Chen:2009zp}. A quick estimate shows that $f_\text{NL}$ becomes unobservably small given the observational constraints on the parameters of the model; in short, the Natural Inflation background requires super-Planckian values of the VEV $v$, which entails that in order to have $M_\text{eff}^2 \approx 4\,\lambda\,v^2 \sim H^2$ we need $\lambda$ to be quite small, implying that $\frac{V_{\sigma\sigma\sigma}}{H} \approx \frac{6\sqrt{2}\,\lambda\,v}{H} \sim \frac{3\sqrt{2}}{2}\frac{H}{v}$ is just too small to produce sizable non-Gaussianities through the use of \eqref{fnlbettest}. This somehow ``negative'' result is at odds with the original naive expectations that through a $(\delta\sigma)^3$ interaction, non-Gaussianities for the adiabatic mode can become large. Although this conclusion is also based on the perturbative assumption that the mixing coupling $\frac{\dot{\theta}_0}{H}$ is small in the QSF regime, we have seen from the single-field EFT point of view that this is indeed the case as we lower down $M_\text{eff}^2$. Even if the single-field EFT does not make sense in the QSF limit, this might shed some light on the real limitations of this particular perturbative condition. Recently there has been renewed interest in non-perturbative (strongly-coupled) QSF models \cite{An:2017hlx,Tong:2017iat,Iyer:2017qzw}. It would be interesting to see if, through these new developments, we could find less supressed signatures of our model. Another venue worth exploring would be to introduce a new scale in the problem, like for example, a non-trivial curvature tensor in field space $\tensor{\mathbb{R}}{^a_b_c_d}$. One way of naturally doing this would be to extend the symmetry group of our model, to a non-abelian one, say $SU(2)$ for definiteness \footnote{As manifolds, $SU(2)$ and the 3-sphere $S^3$ are diffeomorphic, implying that the spectrum of this non-abelian model would consist of three angular (Goldstone) directions plus a radial one. It is not easy to anticipate the phenomenology of such a ``Multi-Quasi-Single-Field'' model.}. All these ideas will be investigated elsewhere.

\acknowledgments
S.R. is grateful to his advisor Zackaria Chacko for initially suggesting this project and having countless discussions, Gonzalo Palma for very useful comments and discussions, and an anonymous referee for suggestions on how to improve the original manuscript. S.R. is supported in part by the National Science Foundation under grant PHY-1620074 and by the Maryland Center for Fundamental Physics. S.R. acknowledges support from Becas Chile and Fulbright-Conicyt scholarships. 

\newpage

\appendix

\section{EFT of the single-field background.}\label{eftofback}

To prove that the single-field EFT of the inflationary background is indeed, to a very good approximation, Natural Inflation, we proceed as follows. As we neglect the dynamics of the radial field, $\sigma$ becomes a Lagrange multiplier, so we can solve algebraically its own equation of motion (which is now a constraint equation) to give $\sigma = \sigma(\theta,\dot{\theta})$. Neglecting time-derivatives of $\sigma$ in \eqref{eqsigma} we find that
\begin{align}\label{constrsigma}
(R + \sigma)\,\dot{\theta}^2 = V_\sigma = (R + \sigma)\left[\lambda\,\{(R + \sigma)^2 - 2\,v^2\} - 2M^2\cos(2\theta)\right],
\end{align}
where we have used \eqref{vsigvthet}. Recalling that $R \equiv +\sqrt{2\,v^2 + \frac{2M^2}{\lambda}}$, it is clear that $\sigma = -R$ is \textit{not} a sensible solution, so \eqref{constrsigma} can be solved for $\sigma$ to give
\begin{align}
\sigma(\theta,\dot{\theta}) = -R + \left(2\,v^2 + \frac{2M^2}{\lambda}\cos(2\theta) + \frac{\dot{\theta}^2}{\lambda}\right)^{1/2}.  
\end{align}
Now we plug this back into the single-field Lagrangian
\begin{align}
\mathscr{L}_{\text{eff}\,\theta} = -\frac{1}{2}(R + \sigma(\theta,\dot{\theta}))^2\tensor{g}{^\mu^\nu}\partial_\mu\theta\partial_\nu\theta - V(\sigma(\theta,\dot{\theta}),\theta),
\end{align}
with $V(\sigma,\theta)$ as given in \eqref{potpol2}. After straightforward algebra one finds that
\begin{align}
\mathscr{L}_{\text{eff}\,\xi} &= \frac{1}{2}\,\dot{\xi}^2 - \widetilde{V}\left(1 - \cos\left(\frac{\xi}{f}\right)\right)\nonumber\\
&+ \frac{\beta}{8(2 + \beta)}\frac{\dot{\xi}^4}{\widetilde{V}} - \frac{\beta}{2(2 + \beta)}\,\dot{\xi}^2\left(1 - \cos\left(\frac{\xi}{f}\right)\right) + \frac{2\beta}{(2 + \beta)}\,\widetilde{V}\sin^4\left(\frac{\xi}{2f}\right),\label{trueeffth}
\end{align}
where $\xi \equiv R\,\theta$, $\beta \equiv \frac{2M^2}{\lambda\,v^2}$, $f \equiv \frac{R}{2}$ and $\widetilde{V} \equiv 4M^2f^2$. The first line in \eqref{trueeffth} reproduces Natural Inflation as given in \eqref{natinf}. Since $\beta \to 0$ as $\frac{m_{\sigma}^2}{H^2} \to \infty$, where $m_{\sigma}^2 \equiv 4\,\lambda\,v^2$ is the ``mass'' of the radial mode that is being integrated out, it is clear that the second line in \eqref{trueeffth} contains operators that are highly suppressed compared to this background theory, so they can be safely neglected, justifying the naive conclusion that the single-field effective background theory is Natural Inflation to a very good approximation.  

\newpage

\section{Effective Field Theory of Inflation.}\label{eftofinf}

A crucial step towards the understanding of the physics of inflation has been the development of the EFT of inflation, where it has been appreciated that the Goldstone degree of freedom associated with the spontaneous breaking of time-translation invariance of time-dependent FLRW backgrounds (such a the quasi-de Sitter background of inflation) can be identified with the inflaton (adiabatic) fluctuation, which is the relevant degree of freedom for the measurements of interest \cite{Cheung:2007st,Creminelli:2006xe}. Indeed, an adiabatic fluctuation is a specific type of perturbation induced by a local, common shift in time $\delta\psi_m(t,\bold{x}) \equiv \psi_m(t + \pi(t, \bold{x})) - \psi_m(t)$, where $\psi_m$ is a set of bosonic matter fields. At linear order, $\delta\psi_m = \dot{\psi}_m\,\pi$. In spatially flat gauge, $\tensor{g}{_i_j} \equiv a^2(t)\,\tensor{\delta}{_i_j}$, all metric perturbations are related to the Goldstone mode by Einstein's equations. For purely adiabatic fluctuations, we can perform a time shift $t \to t - \pi(t,\bold{x})$, to remove all matter fluctuations, $\delta\psi_m \to \delta\psi_m \equiv 0$. This transformation induces an isotropic perturbation to the spatial part of the metric, $\delta\tensor{g}{_i_j} = a^2(t)\,e^{2\,\mathcal{R}(t,\bold{x})}\,\tensor{\delta}{_i_j}$, where $\mathcal{R} = -H\,\pi\,+\,\dots$ and the ellipsis denotes terms that are higher order in $\pi$. In other words, the curvature perturbation $\mathcal{R}$ in comoving gauge is proportional to the Goldstone boson $\pi$ in spatially flat gauge, so for nearly constant $H$ we can think of $\mathcal{R}$ and $\pi$ interchangeably \cite{Baumann:2014nda}. The action for the Goldstone boson can then be constructed by writing down the most general Lorentz-invariant action for the field $U \equiv t\, + \,\pi$ as $S_\pi = \int d^4x\,\sqrt{-g}\,\mathscr{L}\left[U,(\partial_\mu U)^2,\square\,U,\dots\right]$ \cite{Baumann:2011su} (see also \cite{Weinberg:2008hq}). It is striking that the ``universal'' part of the EFT action describes the fluctuations in generic single-field slow-roll inflation models while operators with higher powers in fluctuations with (a priori) arbitrary time-dependent coefficients discriminate between non-slow-roll models. Within this context it is easy to prove that when probing energies beyond the so-called decoupling limit given by $\omega_\text{mix} \equiv \sqrt{\epsilon}\,H$, the analogous of the Goldstone equivalence theorem \cite{Cornwall:1974km} applies, meaning the Goldstone mode $\pi$ decouples from the metric perturbations $\delta\tensor{g}{^\mu^\nu}$ so we can evaluate inflationary models in the unperturbed spacetime. Since for quasi-de Sitter backgrounds $\epsilon \ll 1$, decoupling happens at relatively low frequencies, so the horizon crossing scale $\omega = H$ indeed falls under this regime. We implicitly use this fact throughout our calculations, since corrections to the decoupling limit scenario are of $\mathcal{O}\left(\frac{\omega_\text{mix}^2}{\omega^2}\right)$. This way, we avoid the subleading (however rigorous) unitary gauge Arnowitt-Deser-Misner (ADM) formalism \cite{Arnowitt:1962hi} calculation (for details see \cite{Maldacena:2002vr}). Under these circumstances the action for the Goldstone degree of freedom becomes \cite{Cheung:2007st}
\begin{align} 
S_\pi = \int d^4x\,a^3\bigg(-M_\text{Pl}^2\dot{H}\left[\dot{\pi}^2 - \frac{(\nabla\pi)^2}{a^2}\right] + 2M_2^4\left[\dot{\pi}^2 + \dot{\pi}^3 - \dot{\pi}\frac{(\nabla\pi)^2}{a^2}\right] - \frac{4}{3}M_3^4\,\dot{\pi}^3 + \dots\bigg),\label{pigravityapp}
\end{align}
where $M_2(t)$ and $M_3(t)$ are (a priori) unspecified coefficients of mass dimension $1$ and the $\dots$ stem for higher-order terms in the EFT expansion. Here we note that the coefficient of the time kinetic term $\dot{\pi}^2$ is not completely fixed by the background evolution. In order to avoid instabilities we demand positivity of this coefficient, meaning $-M_\text{Pl}^2\dot{H} + 2M_2^4 > 0$. Since the background spontaneously breaks Lorentz invariance, the speed of sound of $\pi$ waves $c_s^\pi \neq 1$ generically, as $c_s^\pi = 1$ is not protected by any symmetry. From \eqref{pigravityapp} we see that the speed of sound of $\pi$ fluctuations is given by
\begin{align}
(c_s^\pi)^{-2} = 1 - \frac{2M_2^4}{M_\text{Pl}^2\dot{H}},
\end{align}
so the Goldstone action can be rewritten at cubic order as
\begin{align}\label{Goldseffactionapp}
S_\pi = \int d^4x\,a^3\bigg(-\frac{M_\text{Pl}^2\dot{H}}{(c_s^\pi)^2}\left[\dot{\pi}^2 - (c_s^\pi)^2\frac{(\nabla\pi)^2}{a^2}\right] + M_\text{Pl}^2\dot{H}\left(1 - \frac{1}{(c_s^\pi)^2}\right)\left[\dot{\pi}^3 - \dot{\pi}\frac{(\nabla\pi)^2}{a^2}\right] - \frac{4}{3}M_3^4\,\dot{\pi}^3 + \dots\bigg).
\end{align}

\newpage

\section{Non-Gaussianities for the Single-Field EFT.}\label{nongaussdef}

The primary diagnostic for primordial non-Gaussianities is the three-point function or bispectrum which is defined as \footnote{As usual, $\mathcal{R}_\bold{k} \equiv \int d^3x\,\mathcal{R}(\bold{x})\,e^{i\bold{k}\cdot\bold{x}}$.}
\begin{align}
\langle\mathcal{R}_{\bold{p}_1}\mathcal{R}_{\bold{p}_2}\mathcal{R}_{\bold{p}_3}\rangle \equiv (2\pi)^7\delta^{\,3}(\bold{p}_1 + \bold{p}_2 + \bold{p}_3)\,(\Delta_\mathcal{R}^2(p_\star))^2\,\frac{S(p_1,p_2,p_3)}{(p_1\,p_2\,p_3)^2},\label{bispdef}
\end{align}
where $p_\star$ is a fiducial momentum scale, $\Delta_\mathcal{R}^2(k)$ is the (dimensionless) power spectrum given by 
\begin{align}
\langle\mathcal{R}_\bold{k}\mathcal{R}_\bold{k'}\rangle \equiv (2\pi)^5\delta^{\,3}(\bold{k} + \bold{k'})\,\frac{1}{2k^3}\,\Delta_\mathcal{R}^2(k)\label{dimpowspec}
\end{align}
and $S(p_1,p_2,p_3)$ is the shape function which in turn defines a corresponding non-linear parameter $f_\text{NL}$ such that
\begin{align}\label{fnldef}
f_\text{NL} \equiv \frac{10}{9}\,S(p_1 = p_2 = p_3).
\end{align}
There are several shapes that authors have studied thoroughly over the years. One first historical example is the so-called local shape, which is defined through
\begin{align}
S^{\text{local}}(p_1,p_2,p_3) \equiv \frac{3}{10}f_\text{NL}^\text{loc}\left(\frac{p_1^2}{p_2\,p_3} + 2\,\text{perm.}\right)\label{localshape}
\end{align}
and follows from an ansatz (in real space) of the form
\begin{align}
\mathcal{R}(\bold{x}) = \mathcal{R}_g(\bold{x}) + \frac{3}{5}f_\text{NL}^\text{local}\left[\mathcal{R}_g^2(\bold{x}) - \langle\mathcal{R}_g^2\rangle\right],
\end{align}
where $\mathcal{R}_g(\bold{x})$ is a Gaussian random field. In momentum space, the signal peaks for squeezed triangles, $k_1 \ll k_2 \sim k_3$. The local shape arises in models of multifield inflation. On the other hand, in single-field inflation the signal vanishes in the squeezed limit. This is the famous Maldacena's consistency condition \cite{Maldacena:2002vr,Creminelli:2004yq} which reads
\begin{align}
\lim_{k_3 \to 0}\langle\mathcal{R}_{\bold{k}_1}\mathcal{R}_{\bold{k}_2}\mathcal{R}_{\bold{k}_3}\rangle = (2\pi)^3\delta^{\,3}(\bold{k}_1 + \bold{k}_2 + \bold{k}_3)(1 - n_s)\mathcal{P}_\mathcal{R}(k_1)\mathcal{P}_\mathcal{R}(k_3),
\end{align}
where $\mathcal{P}_\mathcal{R}(k) \equiv \frac{2\pi^2}{k^3}\Delta_\mathcal{R}^2(k)$. In other words, for single-field inflation, the squeezed limit of the three-point function is suppressed by $(1 - n_s) \sim \mathcal{O}(\epsilon,\eta)$, so a detection of non-Gaussianities in the squeezed limit can therefore rule out \textit{all} models of single-field inflation.\\
On the other hand, non-Gaussianities associated with a perturbative action like the one in \eqref{Reffaction3} are well known \cite{Chen:2006nt,Chen:2010xka}. In the limit $M_\text{eff} \to \infty$ the bispectrum is of equilateral shape and the contribution from the $\dot{\mathcal{R}}^3$ term gives
\begin{align}
S_{\dot{\pi}^3}(p_1,p_2,p_3) = -6\frac{\dot{\theta}_0^2}{M_\text{eff}^2}\left[1 - 2\frac{\dot{\theta}_0^2}{M_\text{eff}^2}c_s^2\left(1 + \frac{R}{3M_\text{eff}^2}V_{\sigma\sigma\sigma}\right)\right]\frac{p_1\,p_2\,p_3}{(p_1 + p_2 + p_3)^3}.
\end{align}
Then the non-linear parameter is given by
\begin{align}
f_\text{NL}^{\dot{\pi}^3} = -\frac{20}{81}\frac{\dot{\theta}_0^2}{M_\text{eff}^2} + \frac{40}{81}c_s^2\frac{\dot{\theta}_0^4}{M_\text{eff}^4} + \frac{40}{243}\frac{R}{M_\text{eff}^2}V_{\sigma\sigma\sigma}\,c_s^2\frac{\dot{\theta}_0^4}{M_\text{eff}^4},
\end{align}
which is just another way of writting \eqref{fnlpidot3}. It can be shown \cite{Gong:2013sma} that the last term in the above expression is $\lesssim \frac{1}{\eta\,(\Delta_\mathcal{R}^2)^{1/2}}\left(\frac{\dot{\theta}_0^2}{M_\text{eff}^2}\right)$ so it dominates in a ``natural'' model where $\left(\frac{\dot{\theta}_0^2}{M_\text{eff}^2}\right)$ is required to be small, having the chance of giving a non-negligible contribution to $f_\text{NL}$ due to this $(\eta\,(\Delta_\mathcal{R}^2)^{1/2})^{-1}$ prefactor.

\section{Effective Single-Field Theory Regime.}\label{efftheory}

Neglecting $\mathcal{O}(\beta)$ terms as we are dealing with the theory of fluctuations, we can rewrite action \eqref{pertact2} as
\begin{align}
S^{(2)}[g_0,\phi_0,\delta\phi] &= \frac{1}{2}\int d^4x\,a^3\bigg(\left(\frac{\dot{\varphi}_0^2}{H^2}\right)\left\{\dot{\mathcal{R}}^2 - \frac{(\nabla\mathcal{R})^2}{a^2}\right\} + \dot{\mathcal{F}}^2 - \frac{(\nabla\mathcal{F})^2}{a^2} - M_\text{eff}^2\,\mathcal{F}^2 - 4\dot{\theta}_0\left(\frac{\dot{\varphi}_0}{H}\right)\dot{\mathcal{R}}\,\mathcal{F}\bigg),\label{2ndorderfluctaction}
\end{align}
where we have used the fact that $\mathcal{R} \approx -\frac{H}{\dot{\phi}_0}T_a\,\delta\phi^a = -\frac{H}{\dot{\theta}_0}\delta\theta = -\frac{H}{\dot{\varphi}_0}\varphi$ which holds as long as $\dot{\sigma}_0 = 0$ and we have taken $R \approx \sqrt{2}\,v$.\\
Now varying the quadratic action \eqref{2ndorderfluctaction} we get the equations of motion for $\mathcal{R}$ and $\mathcal{F}$ after Fourier transforming spatial coordinates
\begin{align}
\ddot{\mathcal{R}} + (3 + 2\,\epsilon - 2\,\eta_\parallel)\,H\,\dot{\mathcal{R}} + \frac{k^2}{a^2}\,\mathcal{R} &= 2\dot{\theta}_0\left(\frac{H}{\dot{\varphi}_0}\right)\left\{\dot{\mathcal{F}} + \left(3 - \eta_\parallel + \epsilon + \frac{\ddot{\theta}_0}{H\dot{\theta}_0}\right)H\,\mathcal{F}\right\},\label{eqforR}\\
\ddot{\mathcal{F}} + 3H\,\dot{\mathcal{F}} + \frac{k^2}{a^2}\,\mathcal{F} + M_\text{eff}^2\,\mathcal{F} &= -2\dot{\theta}_0\left(\frac{\dot{\varphi}_0}{H}\right)\dot{\mathcal{R}},\label{eqforF}
\end{align}
where $\epsilon \equiv -\frac{\dot{H}}{H^2}$ and $\eta_\parallel \equiv -\frac{\ddot{\varphi_0}}{H\dot{\varphi}_0}$.
As it has been emphasized before $\mathcal{R} = \text{constant}$ and $\mathcal{F} = 0$ are non-trivial solutions to these equations due to the background isometries \cite{Achucarro:2012sm}. Given that $\mathcal{F}$ is heavy, $\mathcal{F}\to 0$ shortly after horizon exit while $\mathcal{R}\to \text{constant}$ as in single field inflation theory. If we consider the short wavelength limit we can neglect the ``Hubble friction'' terms and take $\frac{\dot{\varphi}_0}{H} = \text{constant}$. We also take the physical wave number $p \equiv \frac{k}{a}$ to be a constant in this regime. In this approximation  
\begin{align}
\ddot{\mathcal{R}}_c + p^2\,\mathcal{R}_c &= 2\dot{\theta}_0\,\dot{\mathcal{F}},\\
\ddot{\mathcal{F}} + p^2\,\mathcal{F} + M_\text{eff}^2\,\mathcal{F} &= -2\dot{\theta}_0\,\dot{\mathcal{R}}_c, 
\end{align}
where $\mathcal{R}_c \equiv \left(\frac{\dot{\varphi}_0}{H}\right)\mathcal{R}$. The solutions of this system are given by 
\begin{align}
\mathcal{R}_c &= \mathcal{R}_+\,e^{i\omega_+t} + \mathcal{R}_-\,e^{i\omega_-t},\\
\mathcal{F}_c &= \mathcal{F}_+\,e^{i\omega_+t} + \mathcal{F}_-\,e^{i\omega_-t},
\end{align}
where the frequencies $\omega_\pm$ read \cite{Achucarro:2012yr}
\begin{align}
\omega_\pm^2 =  \frac{M_\text{eff}^2}{2\,c_s^2} + p^2 \pm \frac{M_\text{eff}^2}{2\,c_s^2}\sqrt{1 + \frac{4\,p^2\,(1 - c_s^2)}{M_\text{eff}^2\,c_s^{-2}}}.\label{quadisprel}
\end{align}
Here $(\mathcal{R}_-,\mathcal{F}_-)$ and $(\mathcal{R}_+,\mathcal{F}_+)$ represent the amplitudes of low and high frequency modes respectively and satisfy
\begin{align}
\mathcal{F}_- &= \frac{-2\,i\,\dot{\theta}_0\,\omega_-}{M_\text{eff}^2 + p^2 - \omega_-^2}\,\mathcal{R}_-,\\
\mathcal{R}_+ &= \frac{-2\,i\,\dot{\theta}_0\,\omega_+}{\omega_+^2 - p^2}\,\mathcal{F}_+.
\end{align}
We see that the fields in each pair oscillate coherently.\\
Demanding that the high frequency degrees of freedom do not participate in the dynamics of the adiabatic modes, is only justified in the presence of a hierarchy of the form $\omega_-^2 \ll \omega_+^2$, which is equivalent to demand that $p^2 \ll M_\text{eff}^2\,c_s^{-2}$ by the use of \eqref{quadisprel}. Under these circumstances we get that
\begin{align}
\omega_+^2 &\approx M_\text{eff}^2\,c_s^{-2} \approx M_\text{eff}^2 + 4\,\dot{\theta}_0^2,\label{effcutoff}\\
\omega_-^2 &\approx p^2c_s^2 + (1 - c_s^2)^2\frac{p^4}{M_\text{eff}^2\,c_s^{-2}}.
\end{align}
As far as low energy frequencies are concerned, the condition $p^2 \ll M_\text{eff}^2\,c_s^{-2}$ is equivalent to $\omega_-^2 \ll M_\text{eff}^2\,c_s^{-2}$ so $\omega_+^2 \approx M_\text{eff}^2\,c_s^{-2}$ behaves as the cut-off of the low energy effective theory regime. In this approximation $\mathcal{F}$ is completely determined by $\mathcal{R}_c$ through the relation $\mathcal{F} = \frac{-2\,\dot{\theta}_0\,\dot{\mathcal{R}}_c}{M_\text{eff}^2 + p^2 - \omega_-^2}$. When linear perturbations evolve, their physical wave number $p \equiv \frac{k}{a}$ decreases and the modes enter the long wavelength regime $p^2c_s^2 \lesssim H^2$, where they become strongly influenced by the background and no longer have a simple oscillatory behavior. However, the low energy contributions to $\mathcal{F}$ satisfy $\dot{\mathcal{F}} \sim H\mathcal{F}$ and since we assume $H^2 \ll M_\text{eff}^2$, we can neglect time derivatives in \eqref{eqforF} so we can solve $\mathcal{F}$ in terms of $\mathcal{R}$ as
\begin{align}
\mathcal{F} = -\left(\frac{\dot{\varphi}_0}{H}\right)\frac{2\,\dot{\theta}_0\,\dot{\mathcal{R}}}{\frac{k^2}{a^2} + M_\text{eff}^2}.
\end{align}
Plugging this algebraic relation back into the action \eqref{2ndorderfluctaction} , we get an effective (tree-level) action for the curvature perturbation which at quadratic order reads \footnote{Here $\dbar^3k \equiv \frac{d^3k}{(2\pi)^3}$.}
\begin{align}
S_\text{eff}^{(2)}[g_0,\varphi_0,\mathcal{R}] = \frac{1}{2}\int dt\,\dbar^3k\,a^3\left(\frac{\dot{\varphi}_0^2}{H^2}\right)\left\{\frac{\dot{\mathcal{R}}^2}{c_s^2(k)} + \frac{k^2\,\mathcal{R}^2}{a^2}\right\},
\end{align}
where $c_s^{-2}(k) = 1 + 4\left(\frac{\dot{\theta}_0^2}{\frac{k^2}{a^2} + M_\text{eff}^2}\right)$ is a $k$-dependent speed of sound.

\bibliography{QSFgoldstone}

\end{document}